\documentclass[lettersize,journal]{IEEEtran}
\usepackage{amsmath,amsfonts}
\usepackage{algorithmic}
\usepackage{algorithm}
\usepackage{array}
\usepackage[caption=false,font=normalsize,labelfont=sf,textfont=sf]{subfig}
\usepackage{textcomp}
\usepackage{stfloats}
\usepackage{url}
\usepackage{verbatim}
\usepackage{graphicx}
\usepackage{cite}
\newcommand{\systemName}{\textit{\textit{VizQStudio}}\xspace}

\usepackage{textcomp}
\usepackage{stfloats}
\usepackage{url}
\usepackage{amsmath}
\usepackage{makecell}
\usepackage{balance}
\usepackage{bbm}
\usepackage{amsmath}
\usepackage[hidelinks]{hyperref}
\usepackage{verbatim}
\usepackage{graphicx}
\usepackage{cite}
\usepackage[dvipsnames,svgnames]{xcolor}
\usepackage[most]{tcolorbox}
\usepackage{enumitem}
\usepackage{setspace}
\usepackage{float}
\usepackage{orcidlink}
\usepackage{tabu} 
\usepackage{booktabs}
\usepackage{multirow}
\usepackage[normalem]{ulem}
\useunder{\uline}{\ul}{}
\usepackage{amssymb}
\setlist{noitemsep,parsep=0pt,partopsep=0pt}

\definecolor{review}{HTML}{f72585}
\usepackage{marginnote}
\usepackage{adjustbox}
\global\marginparsep=9pt
\usepackage{mathptmx}                  % use matching math font
\usepackage{CJK}

\definecolor{tvcghighlight}{RGB}{18, 107, 174} 

\definecolor{border}{RGB}{100, 100, 100}

\newtcbox{\mybox}[1][border]
  {on line, arc = 1pt, outer arc = 1pt, colframe = border,
    colback = #1!4!white, boxsep = 0pt, left = 1pt, right = 1pt, top = 1pt, bottom = 1pt, boxrule = 1pt}
    
\AtBeginDocument{

}

\usepackage{tabu}                      % only used for the table example
\usepackage{booktabs}                  % only used for the table example
\usepackage{lipsum}                    % used to generate placeholder text
\usepackage{mwe}                       % used to generate placeholder figures
\usepackage{mathptmx}                  % use matching math font
\usepackage{multirow}
\usepackage{enumitem}
\usepackage{paralist}
\usepackage{balance}

\newcommand{\zixin}[1]{\textcolor{black}{#1}}

\begin{document}

\title{{\systemName}: Iterative Visualization Literacy MCQs Design with Simulated Students}

\author{Zixin Chen\orcidlink{0000-0001-8507-4399},
Yuhang Zeng\orcidlink{0009-0001-7367-2348},
Sicheng Song$^{*}$\orcidlink{0000-0002-2158-0353}, Yanna Lin\orcidlink{0000-0001-9630-9958
}, Xian Xu\orcidlink{0000-0002-2636-7498}, Huamin Qu\orcidlink{0000-0002-3344-9694}
Meng Xia$^{*}$\orcidlink{0000-0002-2676-9032}
        % <-this % stops a space
% \thanks{Zixin Chen, Sicheng Song, Yanna Lin, Huamin Qu are with the Hong Kong University of Science and Technology; Yuhang Zeng is with Carnegie Mellon University; Meng Xia is with Texas A\&M University; 
% Xian Xu is with Lingnan University; 

% }% <-this % stops a space
\thanks{$^{*}$Corresponding author}
% \thanks{This paper was produced by the IEEE Publication Technology Group. They are in Piscataway, NJ.}% <-this % stops a space
\thanks{Manuscript received April 19, 2021; revised August 16, 2021.}}

% The paper headers
\markboth{Journal of \LaTeX\ Class Files,~Vol.~14, No.~8, August~2021}%
{Shell \MakeLowercase{\textit{et al.}}: A Sample Article Using IEEEtran.cls for IEEE Journals}

\IEEEpubid{}
% Remember, if you use this you must call \IEEEpubidadjcol in the second
% column for its text to clear the IEEEpubid mark.

\maketitle

\begin{abstract}
 Multiple-choice questions (MCQs) are a widely used educational tool, particularly in domains such as visualization literacy that require broad conceptual coverage and support diverse real-world applications. 
However, designing high-quality visualization literacy MCQs remains challenging, as instructors must coordinate multimodal elements (e.g., charts, question stems, and distractors), address diverse visualization tasks, and accommodate learners with heterogeneous backgrounds. Existing visualization literacy assessments primarily rely on standardized, fixed item banks, offering limited support for iterative question design that adapts to differences in learners' abilities, backgrounds, and reasoning strategies. To address these challenges, we present \systemName{}, a visual analytics system that supports instructors in iteratively designing and refining visualization literacy MCQs using MLLM-powered simulated students. Instructors can specify diverse student profiles spanning demographics, knowledge levels, and learning-related traits. The system then visualizes how simulated students reason about and respond to different question components, helping instructors explore potential misconceptions, difficulty calibration, and design trade-offs prior to classroom deployment. We investigate \systemName{} through a mixed-method evaluation, including expert interviews, case studies, a classroom deployment, and a large-scale online study. \zixin{Our results indicate that MCQs designed with \systemName{} can support learning outcomes comparable to established benchmark questions, while enabling greater flexibility and scalability during the design process. Overall, this work reframes MLLM-based student simulation in assessment authoring as a design-time, exploratory aid. By examining both its value and limitations in realistic instructional settings, we surface design insights that inform how future systems can support instructor-centered, iterative, and responsible uses of AI for multimodal assessment design in visualization literacy and related domains.}

\end{abstract}

% \begin{abstract}
% This document describes the most common article elements and how to use the IEEEtran class with \LaTeX \ to produce files that are suitable for submission to the IEEE.  IEEEtran can produce conference, journal, and technical note (correspondence) papers with a suitable choice of class options. 
% \end{abstract}

\begin{IEEEkeywords}
Student Simulation, Visualization Literacy Education, Multiple Choice Question Design, MLLM Agents
\end{IEEEkeywords}

\section{Introduction}

As data visualizations increasingly permeate daily life through social media and workplaces, visualization literacy—the ability to accurately interpret and extract insights from visualizations—has become a fundamental skill~\cite{pousman2007casual,segel2010narrative,heer2005vizster,ruckenstein2014visualized,huang2014personal}. With this growing demand, research and educational practices in visualization literacy have expanded rapidly~\cite{stokes2002visual}. 

Effectively teaching and assessing visualization literacy, however, remains challenging due to its broad conceptual scope, diverse reasoning tasks, and varied learner backgrounds. Multiple-choice questions (MCQs) have emerged as a widely used assessment format that aligns well with these needs~\cite{brady2005assessment,butler2018multiple}. Their flexible structure accommodates various contexts and cognitive demands, making them particularly effective for evaluating visualization literacy—from simple information retrieval to higher-level comparative analysis~\cite{lee2016vlat,firat2022interactive,stokes2002visual,alper2017visualization}.

Despite this progress, existing research largely focuses on standardized visualization literacy assessments that use fixed question sets to evaluate proficiency rather than support instruction~\cite{lee2016vlat,ge2023calvi,pandey2023mini}. These tests typically include only one or two questions per knowledge area and are designed for summative benchmarking, offering limited support for adapting questions to diverse learner backgrounds or instructional needs. As a result, educators and students lack sufficient high-quality, diverse, and pedagogically targeted resources for effective teaching and practice, leaving a significant gap in well-vetted MCQ banks for visualization literacy education.

% Moreover, designing high-quality visualization literacy MCQs is inherently challenging and labor-intensive~\cite{collins2006writing,towns2014guide}. Even expert test developers may create only a limited number of well-constructed regular MCQs daily and inadvertently introduce personal biases due to insufficient student feedback~\cite{considine2005design,kim2012incorporation}. The multimodal nature of visualization questions further complicates the design process, as instructors must carefully generate appropriate visual data representations, craft matching question stems and distractors, and address multiple visualization tasks within each question. Additionally, adapting MCQs to the varied skill sets of learners from different fields—such as computer science, business, or design—places substantial burdens on educators, who typically rely on extensive manual effort and experience~\cite{gilbert2005visualization,sorva2013review,presmeg2006research,knaflic2015storytelling}.

Moreover, designing high-quality visualization literacy MCQs is inherently challenging and labor-intensive~\cite{collins2006writing,towns2014guide}. Even expert test developers can produce only a few well-constructed questions per day and may inadvertently introduce personal biases due to limited student feedback~\cite{considine2005design,kim2012incorporation}. The multimodal nature of visualization questions further complicates the process, requiring instructors to generate suitable visual data representations, craft coherent question stems and distractors, and incorporate multiple visualization tasks within each question. Adapting MCQs to learners’ diverse skill sets across fields—such as computer science, business, and design—adds yet another layer of complexity, placing substantial burdens on educators who typically rely on extensive manual effort and expertise~\cite{gilbert2005visualization,sorva2013review,presmeg2006research,knaflic2015storytelling}.

Recent advances in Multimodal Large Language Models (MLLMs) offer promising opportunities for generating educational materials like MCQs. However, their potential remains largely untapped in visualization literacy education. Current research primarily focuses on text-based MCQs~\cite{fu2024qgeval,lamsiyah2024fine,li2024planning,chen2024dr,luo2024chain,xing2024survey} or extract questions from figures~\cite{ding2024exploring}. As a result, most pedagogical considerations crucial to visualization literacy education, like multimodal alignment, diverse reasoning tasks, and learner-specific adaptations, remain inadequately addressed.

To address these challenges, we present \textit{\systemName}, a visual analytics system that supports teachers iteratively create and refine high-quality, multimodal MCQs for visualization literacy education. \textit{\systemName} integrates into instructional workflows, enabling educators to specify sample questions, define visualization tasks, adjust chart features, and configure other question settings. The system also incorporates an MLLM-powered student simulation module, allowing instructors to define diverse student profiles—including demographic attributes, knowledge levels, and cognitive traits—for targeted feedback and comprehensive question evaluations. Educators can interactively adjust question features via intuitive panels, refine details through natural-language prompts, or directly modify specific elements. Meanwhile, instructors receive real-time simulated student feedback enriched by visual analytics that expose underlying reasoning processes. This iterative, student-centered approach helps instructors detect misconceptions, adapt questions to diverse learner needs, and improve overall quality before classroom deployment.

% To address these challenges, we present \textit{\systemName}, a visual analytics system designed to support teachers in iteratively creating and refining high-quality, multimodal MCQs tailored to visualization literacy education. \textit{\systemName} integrates seamlessly into existing instructional workflows, allowing educators to specify sample questions, define visualization tasks, adjust chart features, and manage other question configurations efficiently. Moreover, the system includes a robust student simulation module powered by MLLMs. This module empowers instructors to precisely define diverse student profiles, encompassing demographic attributes, knowledge levels, and cognitive traits, thereby facilitating targeted feedback and comprehensive question evaluations. Educators interactively adjust question features via intuitive panels, utilize natural-language prompts for nuanced refinements, or directly modify specific question elements. Concurrently, instructors receive real-time simulated student feedback enriched by detailed visual analytics that highlight underlying reasoning processes. This iterative, student-centered approach enables instructors to detect potential misconceptions, adapt questions effectively to diverse learner needs, and improve overall question quality before classroom deployment.

Through comprehensive quantitative evaluations of the student simulation and question generation modules, along with illustrative use-case studies, in-depth expert interviews, a real-world classroom experiment, and an online user study, we demonstrate \textit{\systemName}’s capacity to effectively support the design of robust, learner-centered MCQs in visualization literacy education. In summary, our key contributions include:

% \begin{compactitem}
% \item A formative study investigating instructors’ needs in designing visualization literacy MCQs, emphasizing their expectations for simulated student feedback and iterative refinement processes;
% \item \textit{\systemName}, a novel visual analytics system enabling the collaboration between teachers and MLLMs for iterative design of high-quality, multimodal MCQs tailored to diverse visualization literacy educational contexts;
% \item A thorough evaluation highlighting the effectiveness and practical value of \textit{\systemName}, comprising quantitative experiments, expert interviews, illustrative case studies, a classroom deployment and an online user study, underscoring its potential impact on visualization literacy education and broader pedagogical practices.
% \end{compactitem}

\begin{compactitem}
\item \zixin{A formative study investigating instructors' needs in designing visualization literacy MCQs, emphasizing their expectations for simulated student feedback and iterative refinement processes;}
\item \zixin{\textit{\systemName}, a visual analytics system that explores collaboration between teachers and MLLMs for the iterative design of multimodal MCQs across diverse visualization literacy educational contexts;}
\item \zixin{A multi-faceted evaluation that surfaces design insights and practical lessons from deploying \textit{\systemName} in realistic instructional settings, highlighting design insights, practical lessons, and key considerations for visualization literacy education and broader pedagogical practices.}

% underscoring its potential impact and key considerations for visualization literacy education and broader pedagogical practices.}
\end{compactitem}

\section{Related Work}
In this section, we review relevant research, including Visualization Literacy Education, Visual Analytics for Educational Content Authoring, AI-Assisted Multiple-choice Questions
Generation, and LLMs for Student Simulation.
\subsection{Visualization Literacy Education}

Visualization literacy has long been recognized as a crucial topic, particularly given its growing relevance in daily life~\cite{borner2019data,avgerinou1997review}. It is a multifaceted ability involving skills such as reading and interpreting visualizations, extracting information from data, critical reasoning, and authoring effective visualizations. This breadth makes the education of visualization literacy a broad and challenging endeavor~\cite{ruchikachorn2015learning,lo2019learning}.

\zixin
{
Much of the existing literature has focused on summarizing guidelines for effective visual communication~\cite{yang2021explaining,kindlmann2014algebraic,bearfield2024same,nobre2024reading,zhu2022bias} and identifying common pitfalls that hinder comprehension~\cite{lo2023change,lo2022misinformed,hong2025llms}. Other strands of work have introduced standardized visualization literacy assessments, which primarily focus on reading and critical thinking and provide fixed item banks with summative scoring~\cite{lee2016vlat,cui2023adaptive,ge2023calvi}. While these instruments are valuable for benchmarking and summative assessment, their static, fixed-item nature limits adaptability to diverse instructional contexts and learner populations~\cite{firat2022interactive}. 
This limitation is particularly salient in visualization literacy education, where relevant skills have been shown to be multi-dimensional and highly context-dependent rather than uniform or fixed~\cite{beschi2025characterizing}. 
As a result, instructors often require support for formative assessment design that enables them to tailor questions to specific pedagogical goals and learner backgrounds.}

\zixin{Motivated by this need, {\systemName} explores how instructors can be supported in iteratively generating and refining adaptive, multimodal MCQs. 
Rather than replacing existing assessment practices, the system draws on established visualization literacy frameworks and reasoning-oriented assessment perspectives to support instructors' formative question design.
}

\subsection{Visual Analytics for Educational Content Authoring}

\zixin{Visual analytics has been widely used to support educational content authoring by helping instructors inspect and reflect on learner behavior. Prior systems often focus on analyzing students’ problem-solving processes to inform iterative content refinement. 
For example, QLens~\cite{xia2020qlens} and StuGPTViz~\cite{chen2024stugptviz} visualize student reasoning using Sankey diagrams or treemaps, enabling instructors to identify common strategies and misconceptions. 
Other approaches leverage aggregated learner responses to assist content creation, such as summarizing crowdsourced solutions to generate peer-facing hints~\cite{glassman2016learnersourcing}.}

\zixin{Recent work has begun to integrate AI assistance into teacher-driven authoring workflows. 
SPROUT~\cite{liu2024sprout}, for instance, combines interactive visualizations with LLMs to support tutorial creation. 
However, these systems largely rely on instructors’ expertise or historical learner data to guide refinement. 
They offer limited support for exploring how diverse learners might respond to newly authored content.}

\zixin{This reliance on previously collected data presents a fundamental limitation: for novel questions or instructional materials, such data is unavailable. 
As a result, instructors lack tools to proactively reason about potential learner responses during the design phase. 
{\systemName} addresses this gap by using simulated student profiles to visualize plausible reasoning paths, enabling instructors to anticipate misconceptions and iteratively refine questions prior to classroom use.}

\subsection{AI-Assisted Multiple-choice Questions Generation} 

Multiple-choice questions are widely used in education~\cite{little2015optimizing}, yet designing high-quality MCQs remains notoriously challenging~\cite{collins2006writing,towns2014guide}. Even experienced test developers often produce only a few well-crafted questions per day~\cite{kim2012incorporation}, and creating isomorphic problems with varied distractors becomes especially difficult when images and figures are involved. Recent advances in generative AI, particularly Large Language Models, have opened new avenues for assisting teachers in designing MCQs. Prior studies have focused on automatically generating information-extraction questions from text passages~\cite{fu2024qgeval,li2024planning,yuan2022selecting,zeng2024advancing}, refining distractor design, or supporting varied question difficulties~\cite{chen2024dr,luo2024chain}.

However, these approaches often lack the diverse pedagogical coverage needed for real-world visualization literacy education, including comprehensive knowledge coverage, student diversity, and domain-specific requirements such as visualization accessibility. While some tools leverage human-AI collaboration to improve question design with pedagogical considerations, they are generally limited to open-ended formats~\cite{wang2019upgrade, lu2023readingquizmaker}, subjects such as coding~\cite{sarsa2022automatic}, or peer-generated items used for learning~\cite{yeckehzaare2020qmaps}. In contrast, our work leverages MLLMs to specifically support the design of adaptive, multimodal MCQs for visualization literacy, ensuring coverage across diverse knowledge points while accommodating learners of varying backgrounds and skill levels.

\subsection{LLMs for Student Simulation}

\zixin{
Recent research has explored the use of LLMs to simulate student behavior across a range of educational contexts. 
Multi-agent systems have been proposed to model classroom interactions or virtual learning environments~\cite{thomas2023so,zhang2024simulating,xu2025classroom}, while other work focuses on supporting teacher training through simulated tutoring, problem-solving scenarios, or dialog-based interactions~\cite{lee2023generative,pan2025tutorup,markel2023gpteach,jin2024teachtune}. 
Collectively, these studies demonstrate the feasibility of LLMs as student surrogates.
However, they primarily address pedagogical interaction during instruction, rather than providing \emph{design-time feedback} for assessment authoring—where no empirical student data exists, and instructors must reason about how diverse learners may respond to newly created questions before deployment.
}

% \zixin{Parallel efforts have enriched student simulation by incorporating cognitive factors (e.g., knowledge level) and non-cognitive traits (e.g., personality or mindset)~\cite{choi2024proxona,chuang2023simulating,he2024evaluating,lu2024generative}. 
% While some work models these traits at finer granularity~\cite{liu2024personality}, such simulations are often domain-agnostic and offer limited insight into how different learners may reason through specific question content, particularly in visualization-intensive tasks.}

\zixin{
Parallel efforts have enriched student simulation by incorporating cognitive factors (e.g., knowledge level) and non-cognitive traits (e.g., personality or mindset)~\cite{choi2024proxona,chuang2023simulating,he2024evaluating,lu2024generative}. 
While such domain-agnostic models can be applied across tasks, they often provide limited diagnostic value for assessment design in visualization-intensive domains. 
Visualization literacy relies on domain-specific reasoning bottlenecks—such as visual encoding interpretation, proportional comparison, and perceptual shortcuts—that are not explicitly captured in general-purpose simulations. 
As a result, these approaches offer limited insight into how diverse learners may reason through specific visualization questions, constraining their usefulness for iterative assessment refinement.
}

\zixin{Our work is situated at this intersection, but addresses a different design goal. 
Rather than using student simulation for instruction or teacher training, we explore its role as a design-time instrument for assessment authoring. 
By grounding LLM-based agents in domain-specific visualization literacy factors (e.g., visual processing ability, chart interpretation strategies) and visualizing their simulated reasoning, {\systemName} supports instructors in anticipating misconceptions and iteratively refining MCQs before deployment. 
In this framing, student simulation complements empirical learner data by providing exploratory feedback when real responses are unavailable.}

\section{Formative Study}
\label{sec:formative_study}
To better understand the challenges instructors face in designing visualization literacy questions, as well as their needs for a tool involving simulated students, we conducted semi-structured interviews with six domain experts (E1–E6). The group comprised three assistant professors (E1–E3), one lecturer (E4), and two postdoctoral researchers (E5 \& E6), each with at least three years of experience teaching or assisting in undergraduate and master's data visualization courses, with extensive experience in homework and exam design. During the interviews, we explored experts’ question design workflows, prior experiences with AI tools (e.g., LLMs), expectations for student simulation, and additional functional needs. Below, we summarize the key design requirements that emerged from these discussions.

% \ToDo{Directly add the expected analysis requirement for simulated student results analysis? => intro focus more on the question evaluation and student simulation?}

% 1. input format 2. Knowledge points & chart focus 3. configurations like difficulty/ scenarios

\textbf{R1: Flexible Sample Question and Requirement Specification.} Instructors often begin the question-design process with varying levels of readiness. Some arrive with fully developed samples, others bring only raw screenshots from course slides or pre-made visualizations, and still others possess little more than a broad idea. Consequently, the system must accommodate these diverse entry points, enabling teachers to either build on existing materials or generate new content from scratch. Moreover, it should offer flexible configuration options—such as adjusting question difficulty, specifying contextual details (e.g., domain scenarios), choosing chart types, and identifying core knowledge points—so that instructors can tailor questions to their specific educational goals and the diverse needs of their students.

\begin{figure}[!h]
    \centering
    \includegraphics[width=0.9\linewidth]{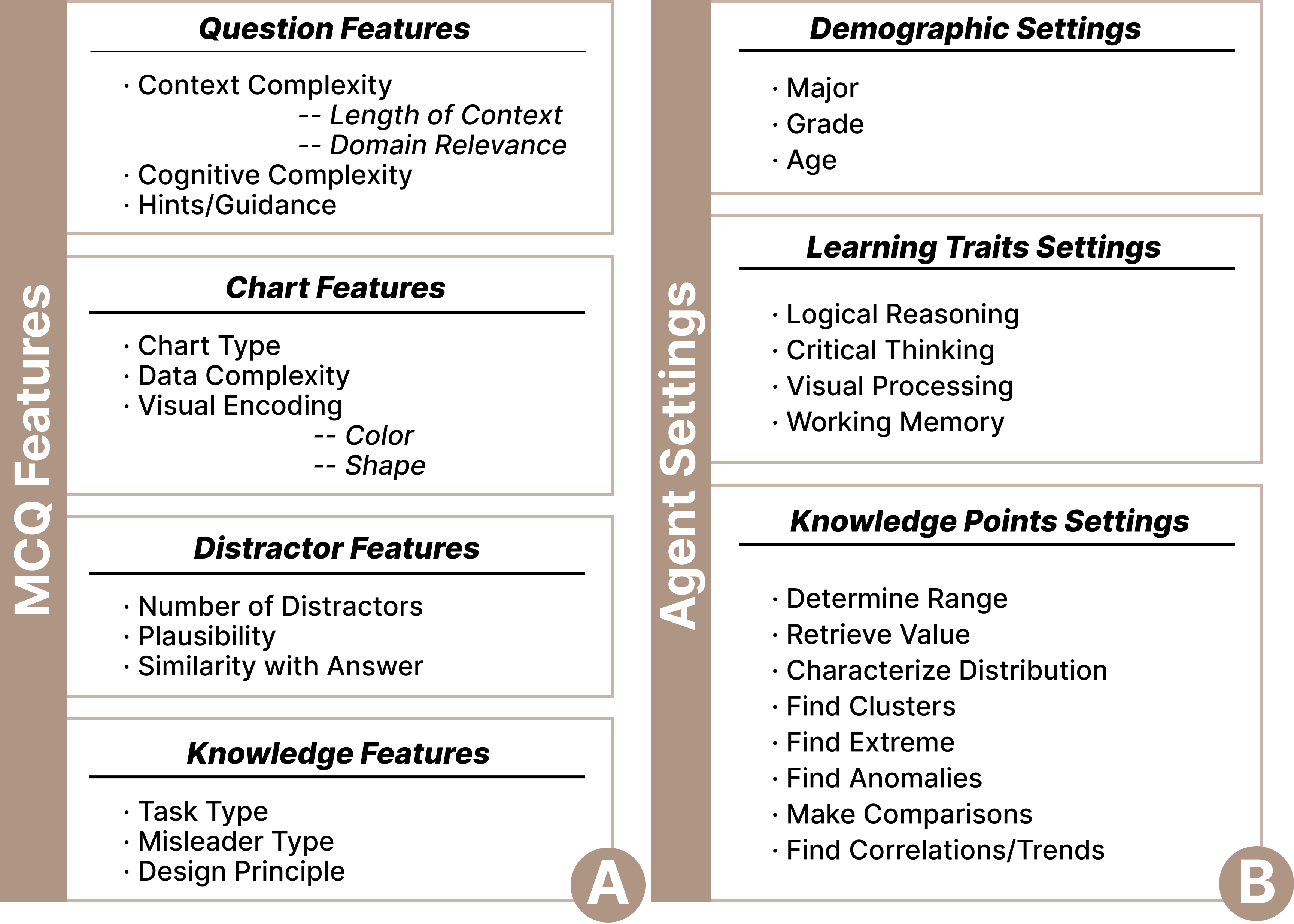}
    \caption{ Overview of the proposed MCQ features and student agent profile settings, with detailed definitions provided in the supplementary. \textbf{(A)} MCQ features are organized into four categories: question, chart, distractor, and knowledge. \textbf{(B)} Agent settings cover demographic attributes, learning traits, and knowledge point relevant to visualization literacy.}
    \label{fig:Feature_Details}
\end{figure}

To further inform these configuration needs, we conducted follow-up interviews with each expert. We presented sample MCQs from VLAT~\cite{lee2016vlat} and CALVI~\cite{ge2023calvi}, asking which adjustable settings and student profile attributes would be most beneficial. \zixin{To reduce subjective bias and strengthen theoretical grounding, we triangulated experts' feedback with established educational and visualization literacy frameworks. 
This process resulted in four categories of MCQ features (14 in total), as summarized in~\autoref{fig:Feature_Details}-A. 
Specifically, we aligned \textit{Cognitive Complexity} with the \textit{Revised Bloom’s Taxonomy}~\cite{KrathwohlDavidR.2002ARoB} to capture a progression from lower-order tasks (e.g., value retrieval) to higher-order analytical reasoning. 
In parallel, content-specific features (e.g., visual encoding and chart interpretation) were grounded in validated visualization literacy frameworks~\cite{lee2016vlat, ge2023calvi}, ensuring that question design reflects established constructs in visualization education.}

\textbf{R2: Diverse Student Profiles Considerations.} 
Participants emphasized the importance of using LLM-driven simulations to capture the breadth of student diversity when designing and evaluating questions. They highlighted the need to represent a wide range of ages, backgrounds, knowledge levels, and personal traits. \zixin{To operationalize these considerations in a theory-informed manner, we triangulated expert input with prior work in learning sciences and visualization cognition. 
This resulted in three categories of 15 student profile features—demographic attributes, learning traits, and visualization-relevant knowledge points—as summarized in~\autoref{fig:Feature_Details}-B.} 

\zixin{In particular, participants stressed factors specific to visualization literacy, such as visual processing abilities and working memory. 
This emphasis is grounded in Cognitive Load Theory~\cite{SwellerJohn1988Cldp} and models of graph comprehension~\cite{CarpenterPatriciaA1998AMot}, which identify these faculties as key constraints when learners decode visual encodings while reasoning about data relationships. By incorporating these traits into student profiles, the system is designed to better reflect variations in learners’ prior exposure and cognitive demands, providing instructors with a structured lens to reason about novice-to-advanced differences during question design.}

In addition, participants noted that simulated students should provide reasoning traces for their answers, enabling instructors to identify misconceptions and refine question design. Finally, for scenarios where instructors lack detailed insights into the student population, they recommended offering a set of default student profiles. These profiles would help evaluate how well questions cater to general learners without requiring customized input, thereby promoting fairness and broader applicability in assessment design.

\textbf{R3: Concise Summaries for Simulation Results with Details on Demand.} A key advantage of using MLLMs for question generation and student simulation is the ability to produce diverse question settings and predict how learners might respond. However, instructors can be overwhelmed by the volume of generated questions and simulation data. Given the complexity of designing each high-quality MCQ, participants recommended focusing on one question at a time and providing concise summaries of student feedback. They suggested grouping similar students to present aggregated results while highlighting critical metrics such as predicted difficulty, error rates, and student perceptions to help instructors quickly assess question quality and identify refinements. 

Additionally, participants valued on-demand, granular analyses of students' reasoning. The ability to ``drill down'' into individual thought processes for specific answers helps uncover misconceptions and confirm alignment with intended designs. They recommended a hierarchical feedback approach: start with high-level insights and allow deeper exploration into detailed reasoning as needed.

\textbf{R4: Iterative Question Design and Refinement.} A core requirement for creating high-quality MCQs that effectively incorporate simulated student feedback is to support continuous teacher-driven refinement of all question components (e.g., the stem, options, and associated visualizations). During interviews, participants emphasized the need for straightforward methods to make adjustments, either by describing desired revisions to the MLLM or by directly manipulating settings via simple buttons (e.g., chart type, dataset complexity, color schemes or question context). After each update, the system should re-simulate student responses, allowing instructors to compare performance before and after modifications. This feedback loop enables teachers to iteratively refine MCQs, ensuring robust pedagogical rigor.

\begin{figure*}[t]
    \centering
    \includegraphics[width=\textwidth]{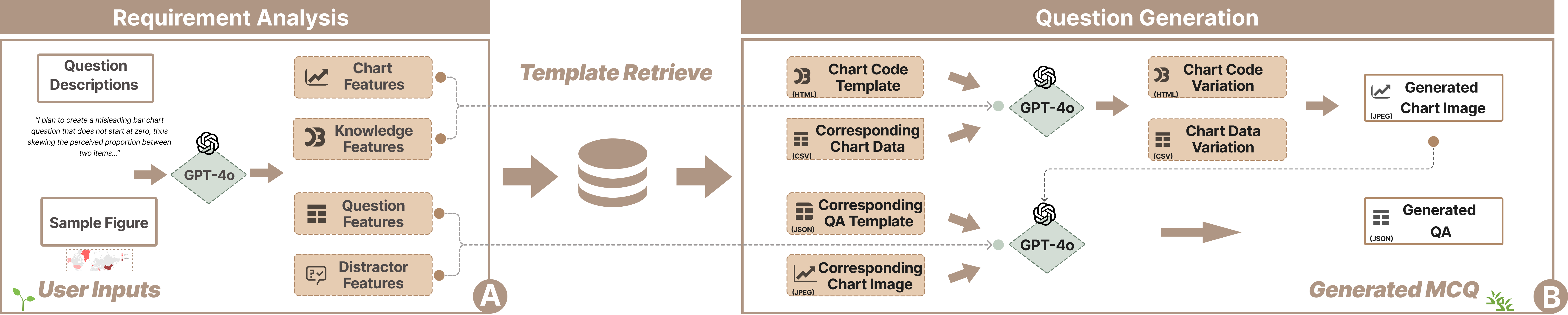}
    \caption{ The question generation pipeline. \textbf{(A)} The system analyzes the instructor's input—such as textual descriptions, sample figures, or lecture slides—using a GPT-4o module. It then extracts feature requirements (e.g., chart features, knowledge points, distractor features) in accordance with the feature set described in~\autoref{fig:Feature_Details} \textbf{(B)} These extracted features are used to retrieve the most similar MCQ templates, along with corresponding chart data and code, from an external database we collected. \textbf{(C)} In the first step, the system uses the retrieved D3.js code, data, and QA templates—combined with the corresponding chart image and user-specified feature requirements—to modify the template code and produce a customized chart figure. In the second step, an additional MLLM module processes the generated chart image, the template QA, and the instructor's requirements for question content and distractors to generate the final QA set. This two-step process yields an output MCQ that fully aligns with the instructor's design intent.}
    \label{fig:QuestionGeneration}
\end{figure*}

\textbf{R5: Balance Difficulty Across the Question Set.} Instructors emphasized the importance of designing not just individual questions, but a well-balanced set with varied difficulty levels. A diverse and calibrated question set better supports learners' progression and ensures comprehensive coverage of visualization literacy skills.

\section{Question Generation and Student Simulation}
\label{sec:question_generation_simulation}
This section details the MLLM-based question generation and student simulation modules, which form the backbone of our system’s iterative design workflow.

% In this section, we detail the design process behind the MLLM-based question generation and student simulation modules, both of which are essential to our system's iterative design workflow.

% As shown in Figure~XX, the overall process comprises five sequential steps: (A) \emph{Requirement Collection}, (B) \emph{Initial Question Generation}, (C) \emph{Adjustable Feature Integration}, (D) \emph{Student Simulation}, and (E) \emph{Question Evaluation \& Refinement}. We first introduce our Teacher-MLLM collaborative approach to question generation (Section~\ref{subsec:teacher_mllm_generation}), followed by an explanation of our MLLM-based student simulation pipeline (Section~\ref{subsec:mllm_simulation}), which provides rapid feedback and performance predictions for iterative question refinement.

\subsection{Teacher-MLLM Collaborative Question Generation}
\label{subsec:teacher_mllm_generation}
Designing high-quality MCQs for visualization literacy requires generating accurate, diverse, and sometimes intentionally misleading charts — a significant technical challenge. We first outline our technical stack choices and the retrieval-augmented generation (RAG) approach used to improve MLLM-generated chart quality, followed by the backend pipeline for question generation and iterative refinement.

% Generating high-quality MCQs for visualization literacy hinges on producing accurate and compelling chart visualizations, which is a significant technical challenge. In this section, we first describe our initial investigation and the technical stack choices for visualization generation, along with the preparation of a reference dataset used in a retrieval-augmented generation (RAG) approach to enhance the quality of MLLM-generated charts. We then detail our backend pipeline for question generation and iterative refinement.

% To generate high quality MCQs, the biggest technical challenges lays on generate high quality chart visualization. Thus, in the following section, we first introduce our initial investigation and technical stack decisions for visualization generation, together with our reference dataset preparation for a RAG method to enhance MLLM chart generation quality, then we details our question generation and iterative refinement backend pipeline. 

\subsubsection{Visualization Generation Method Selection}
The system should transform varied user inputs, from rough descriptions and lecture slides to fully formed samples, into new MCQs (\textbf{R1}). A key challenge is generating precise chart visualizations, as current MLLMs remain unstable~\cite{zeng2024advancing}. While prior work often uses Vega-Lite, its high-level grammar lacks the flexibility needed for tasks involving misleaders (e.g., truncated axes, non-linear scales). Instead, we leverage MLLMs to generate D3.js code, which offers fine-grained control and supports complex visualization designs~\cite{chen2025unmasking}.

\begin{figure}[!bp]
    \centering
    \includegraphics[width=\linewidth]{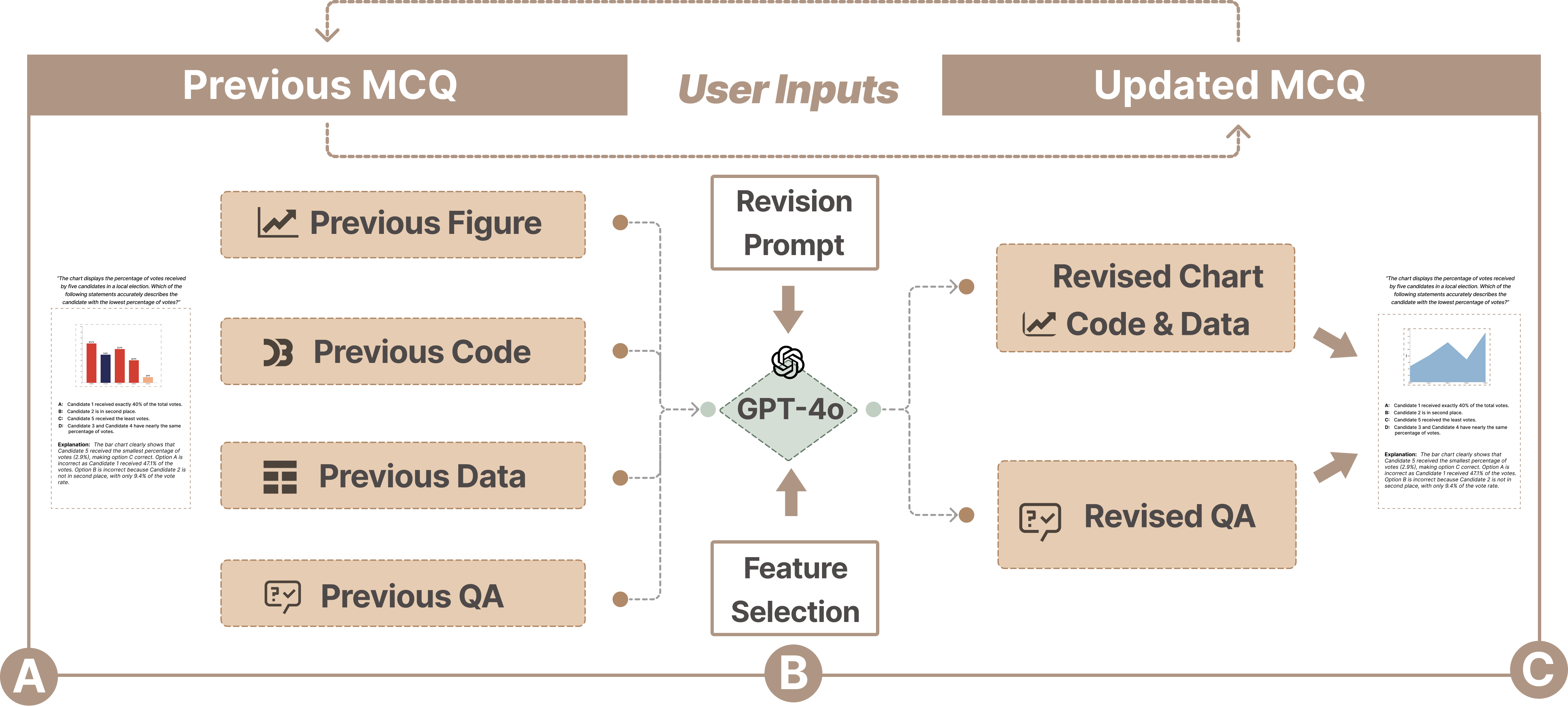}
    \caption{ Overview of the iterative MCQ update process. (A) The system starts with the previous MCQ (chart image, code, data, and QA content) as references. (B) Alongside these references, the MLLM module also takes in the user’s revision prompt and selected features from the \textbf{Feature View}. (C) The MLLM then updates the chart code and CSV data, renders the revised chart, and modifies the QA content, producing an updated MCQ aligned with the user's instructions.}
    \label{fig:Iterative_Update}
\end{figure}

% \subsubsection{Code Templates Collection, Question Generation and Iterative Refinement}
\subsubsection{Template-Based Generation and Iterative Refinement}
Directly generating D3.js from scratch often produces unstable scripts~\cite{chen2025unmasking}. To improve reliability, we adopt a retrieval-augmented generation (RAG)-inspired approach: instead of creating visualizations entirely from scratch, the system queries, edits, and varies existing templates. To achieve this, we curated 80 standardized D3.js templates covering ten common chart types and all misleaders from existing visualization literacy tests~\cite{lee2016vlat,pandey2023mini,ge2023calvi}.

% Directly generating D3.js from scratch often leads to unstable scripts~\cite{chen2025unmasking}. To improve reliability, we adopt a similar approach to retrieval-augmented generation (RAG): instead of creating entire visualizations, the system query, edits and varies existing templates. We curated 80 standardized D3.js templates covering ten common chart types and all misleaders from existing visualization literacy tests~\cite{lee2016vlat,pandey2023mini} and CALVI~\cite{ge2023calvi}.

% One significant drawback of using MLLMs for generating D3.js code is that they often produce scripts that lack stability and precision, necessitating further revisions to render correctly in the browser~\cite{chen2025unmasking}. To mitigate this issue, instead of letting the model create entire visualizations from scratch, we borrow the idea of RAG~\cite{gao2023retrieval} to utilize MLLMs edit and vary existing D3.js chart code templates. Specifically, we developed standardized D3.js templates for ten common chart types found in existing visualization literacy tests~\cite{lee2016vlat,pandey2023mini}. Additionally, we leveraged an open-source dataset that reproduces misleading visualization QAs from the CALVI question set in D3.js, yielding approximately 80 code templates~\cite{chen2025unmasking}. Together, these templates cover most of the chart types used in current visualization literacy tests. 

As shown in \autoref{fig:QuestionGeneration}, we developed a two-stage question generation pipeline with an iterative refinement process. First, the requirement analysis module parses the user’s textual or image input to extract four types of MCQ features (\autoref{fig:Feature_Details}-A). Based on these features, the system retrieves the most relevant D3.js template and CSV data. The question generation module then modifies them according to user configurations, introducing variations (e.g., random colors or data distributions) to enrich the resulting charts (\autoref{fig:Feature_Details}-B). This pipeline reliably generates chart visualizations that meet the diverse needs of visualization literacy MCQs, as further evaluated through quantitative experiments in \autoref{sec:evaluation}.

% Leveraging these code templates, we developed a two-stage question generation pipeline alongside an iterative refinement process. As illustrated in~\autoref{fig:QuestionGeneration}, the pipeline first employs a requirement analysis module to parse the user's textual or image input and extract the four types of MCQ features (\autoref{fig:Feature_Details}-A). With the extracted features, the system retrieves the most relevant D3.js template along with the corresponding CSV data. Question generation module then modifies them according to the user’s configuration choices, introducing variations (e.g., random color choices or data distributions) to enrich the resulting chart (\autoref{fig:Feature_Details}-B). Consequently, our system reliably generates chart visualizations that meet the diverse requirements of visualization literacy MCQs. We evaluate the speed and quality of visualizations generation through quantitative experiments, detailed in \autoref{sec:evaluation}.

Furthermore, \autoref{fig:Iterative_Update} illustrates the iterative refinement process: instructors can update any part of an MCQ (chart, code, data, or QAs) via revision prompts and feature selections. The system integrates these changes using MLLMs to produce updated versions, enabling instructors to progressively improve question quality and pedagogical rigor.

\subsection{MLLM-Based Student Simulation}
\label{subsec:student_simulation}
% To obtain precise feedback and evaluation for the generated MCQs, and to provide actionable revision suggestions for instructors, we employed an MLLM-based student agent simulation component. This module was developed based on instructors' requirements and expectations \textbf{(R2)} and incorporates the controllable features outlined in~\autoref{fig:Feature_Details}. In the following section, we detail the implementation of our student simulation modules and the associated question feedback mechanism.
To provide precise feedback on generated MCQs and actionable revision suggestions, we implemented an MLLM-based student simulation module. Designed based on requirements \textbf{(R2)}, it integrates the controllable features in~\autoref{fig:Feature_Details} and supports three core processes: generating student profiles, clustering students, and simulating students responses. \zixin{Unlike Item Response Theory (IRT)\cite{KinknerTamra1993Foir} or Bayesian Knowledge Tracing (BKT)\cite{CorbettAlbertT.1995KtMt}, which depend on historical learner response data, our approach instantiates synthetic learner personas grounded in User Modeling frameworks~\cite{RichElaine1979UMvS, KobsaAlfred2001GUMS}. This design enables formative, design-time feedback for newly authored questions without requiring prior learner interaction data.}

\textbf{Generating and Updating Student Profiles:} 
A core requirement is representing a diverse range of students \textbf{(R2)}. We developed a profile generation module that dynamically synthesizes profiles using GPT-o3-mini, allowing fine-grained control over demographic, cognitive, and knowledge attributes (\autoref{fig:Feature_Details}-B). \zixin{The system honors instructor-defined constraints and otherwise applies sensible defaults. For example, the default simulated cohort size is set to 20, reflecting the typical size of a university tutorial or discussion group~\cite{exley2004small}. This choice provides a reasonably representative snapshot of learner diversity while maintaining computational efficiency for real-time interaction. Instructors can override this default by specifying any cohort size. Prompt templates were iteratively refined for accurate profile construction, with full templates provided in the supplementary materials.}

\textbf{Clustering Students for Aggregated Analysis:} Analyzing individual responses at scale can be overwhelming, so our system clusters students based on three levels of features (~\autoref{fig:Feature_Details}-B) using K-means (default $K=4$), a common approach in educational contexts~\cite{sharif2024application, arslan2023clustering}. The results are visualized with radar charts to summarize group characteristics and support cross-group comparisons (\textbf{R3}). This enables instructors to efficiently assess group-level performance and identify groups struggling with specific question types.

\textbf{Simulating Student Responses:} 
To evaluate MCQ effectiveness, we simulate student responses using an MLLM-based reasoning framework conditioned on each agent’s demographic attributes, cognitive traits, and domain knowledge. This enables exploration of how different students answer the question, supporting analyses of answer distributions, potential misconceptions, and subsequent question refinements.

The simulation pipeline retrieves the student profile, the target MCQ (stem and options), and a base64-encoded chart image to construct a structured reasoning prompt. Each student’s cognitive abilities (e.g., logical reasoning, visual processing) are mapped to selection likelihoods for different reasoning pathways. For each question, the simulated student produces: (i) a selected answer; (ii) a step-by-step reasoning trace; and (iii) ratings of the question along six dimensions—context clarity, chart complexity, data difficulty, visual encoding complexity, overall cognitive challenge, and hint dependency.

\zixin{It is important to note that all of these outputs are generatively inferred by the MLLM. As no prior student data exists for novel MCQs, the model is prompted to ``think aloud'' step by step while adhering to the constraints specified in the student profile and knowledge configuration. The resulting reasoning traces are therefore not treated as ground-truth explanations of individual learner cognition, but as plausible reasoning pathways whose instructional relevance is examined empirically through alignment with real students' self-reported reasoning in our large-scale online study (\autoref{sec:evaluation}).} 

These outputs support both aggregate summaries and fine-grained diagnostics of question quality, enabling instructors to inspect how different learner profiles may reason through a question and where misconceptions or unintended shortcuts may arise. For efficiency, simulations are parallelized via multi-threaded API calls to reduce latency in large-scale runs. The outputs are parsed to extract answers, reasoning traces, justifications, and diagnostic metrics, supporting iterative MCQ refinement while maintaining alignment with diverse learner profiles.

\section{System Design}
\label{sec:system_design}

Based on the design requirements identified in \autoref{sec:formative_study} and modules in \autoref{sec:question_generation_simulation}, we developed \textit{{\systemName}}, a visual analytics system that enables instructors to iteratively design and refine visualization literacy MCQs using MLLM-based simulated student feedback. This section first outlines the system workflow, then introduces the four main input panels (\textbf{A}, \textbf{B}, \textbf{D}, \textbf{E}), and finally presents the Simulation View (\textbf{C}), which displays the generated questions and simulation results.

% \begin{figure}[!tbp]
%     \centering
%     \includegraphics[width=\linewidth]{figs/Teaser_v7.png}
%     \caption{System interface of {\systemName}. The interface is organized into five main panels: (A) \textbf{Sample View}, where users provide a sample question and/or reference figure; (B) \textbf{Feature View}, which displays four feature categories and historical user choices to control and adjust question generation; (C) \textbf{Simulation View}, which classifies the simulated students and presents both the generated question and simulation outcomes; (D) \textbf{Agent Profile View} where users broadly define the characteristics of the student agents; and (E) \textbf{Agent Setting View}, which enables fine-grained control and adjustment of the MLLM-simulated students' profiles. The four side panels (A, B, D, E) serve as user control areas, while the central (C) \textbf{Simulation View} aligns each component of the \textit{Generated Question} with its corresponding simulated feedback in \textit{Simulation Results} zone. This layout allows users to immediately see how each round of edits or feature adjustments influences student responses. 
%     }
%     \label{fig:teaser}
% \end{figure}

\begin{figure*}[ht]
\centering
\includegraphics[width=\linewidth]{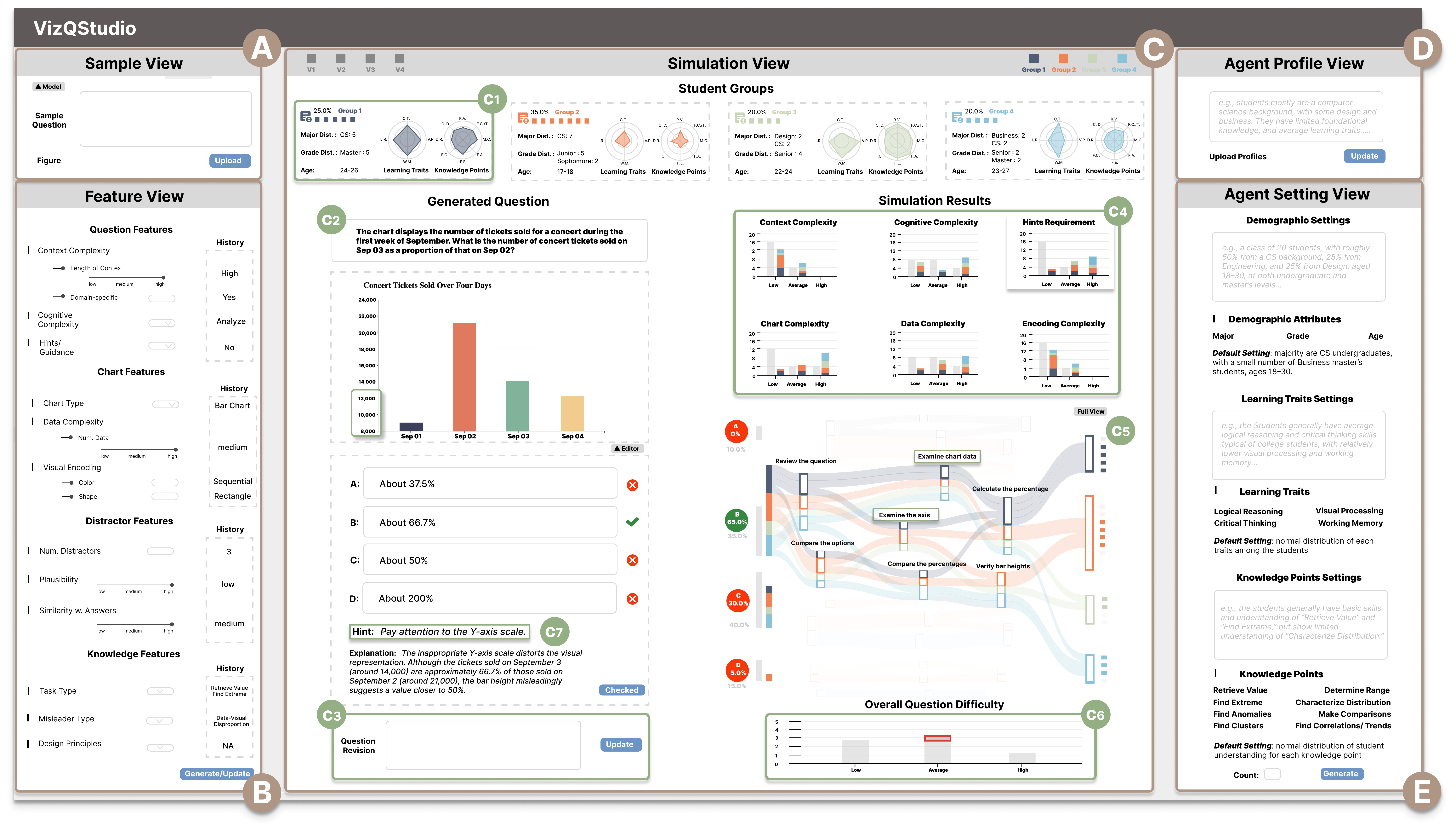}   
\caption{System interface of {\systemName}. The interface is organized into five main panels: (A) \textbf{Sample View}, where users provide a sample question and/or reference figure; (B) \textbf{Feature View}, which displays four feature categories and historical user choices to control and adjust question generation; (C) \textbf{Simulation View}, which classifies the simulated students and presents both the generated question and simulation outcomes; (D) \textbf{Agent Profile View} where users broadly define the characteristics of the student agents; and (E) \textbf{Agent Setting View}, which enables fine-grained control and adjustment of the MLLM-simulated students' profiles. The four side panels (A, B, D, E) serve as user control areas, while the central (C) \textbf{Simulation View} aligns each component of the \textit{Generated Question} with its corresponding simulated feedback in \textit{Simulation Results} zone. This layout allows users to immediately see how each round of edits or feature adjustments influences student responses. 
}
\label{fig:teaser}
\end{figure*}

\subsection{System Overview}
\label{sec:system_overview}

% \textit{{\systemName}} follows an iterative workflow that helps teachers design, test, and refine MCQs for visualization literacy by leveraging rich feedback from simulated students. The process begins with \textit{Question Drafting}, where teachers describe key points or paste an initial sample question stem and its accompanying chart. Next, in \textit{Feature Adjustment and Profile Customization}, teachers adjust question features (e.g., chart type, knowledge points) and define the characteristics of the simulated students (e.g., background, learning traits). Once these elements are set, the \textit{Simulation and Review} step uses the defined student profiles to provide immediate feedback and response to the newly generated or revised question, revealing potential shortcomings or biases in the question design. Finally, teachers \textit{Refine} the MCQs based on this feedback, iterating until the questions align with their educational goals. By repeating this cycle, teachers can create MCQs tailored to diverse learner needs and skill levels, ensuring that each question is both pedagogically sound and effectively targeted.

\textit{{\systemName}} follows an iterative workflow that helps instructors design, test, and refine visualization literacy MCQs using MLLM-based simulated student feedback.
Teachers start by drafting a question or importing sample slides, then adjust question features (e.g., chart type, knowledge points) and configure student profiles (e.g., background, learning traits). The system then simulates student responses and reasoning, providing immediate insights into potential misconceptions, and question quality. Instructors refine the questions based on this feedback and re-run simulations until the MCQs align with instructional goals and accommodate diverse learner needs.

\begin{figure}[!tbp]
    \centering
    \includegraphics[width=\linewidth]{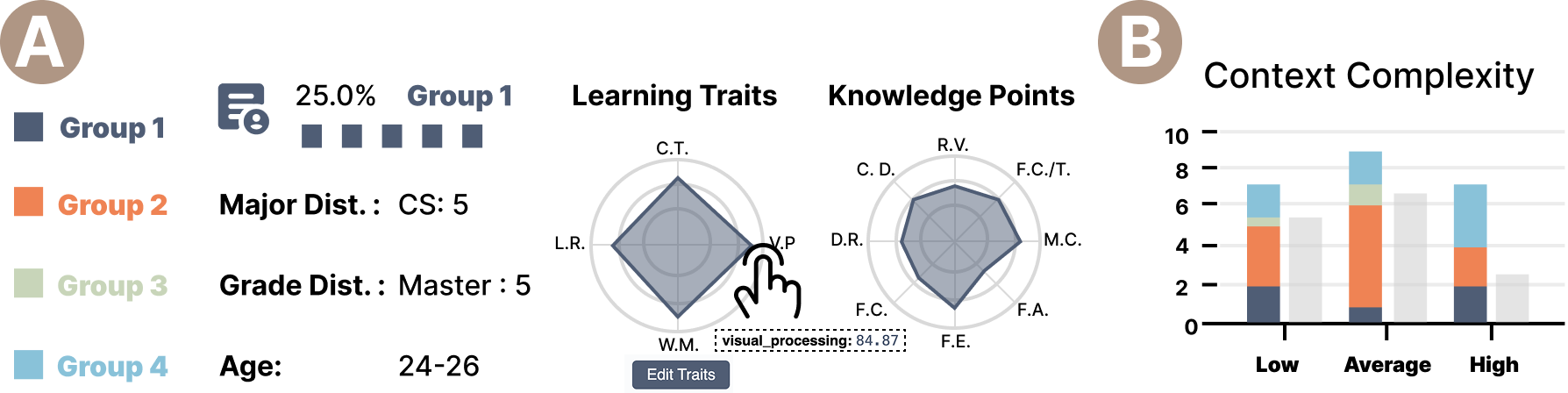}
    \caption{(A) Each color-coded group card displays the cluster size and demographic details. Radar charts illustrate each group's average traits and knowledge levels. Users can interactively drag the radar charts to adjust the group levels. (B) Each stacked bar chart illustrates each student group's feedback on the question stem and related features. Colored bars represent feedback for the current version, while grey bars indicate responses from the previous version for quick comparison. 
    % By comparing these bars, instructors can quickly assess how changes in the question design or features influence perceived clarity and difficulty across different student groups.
    }
    \label{fig:groupCard}
\end{figure}

% \begin{figure}[!tbp]
%     \centering
%     \includegraphics[width=\linewidth]{figs/chart_edit.png}
%     \caption{\revision{Users can expand the chart editor panel in (A) to review the original CSV data (B) and D3.js/HTML code (C), and edit them directly.} 
%     % By comparing these bars, instructors can quickly assess how changes in the question design or features influence perceived clarity and difficulty across different student groups.
%     }
%     \label{fig:chartEdit}
% \end{figure}

\subsection{User Interface}
\label{sec:system_ui}

The interface of \textit{\systemName} is organized into five main panels (~\autoref{fig:teaser}), each corresponding to a distinct stage in the question design process.

\begin{figure*}[t]
    \centering
    \includegraphics[width=\textwidth]{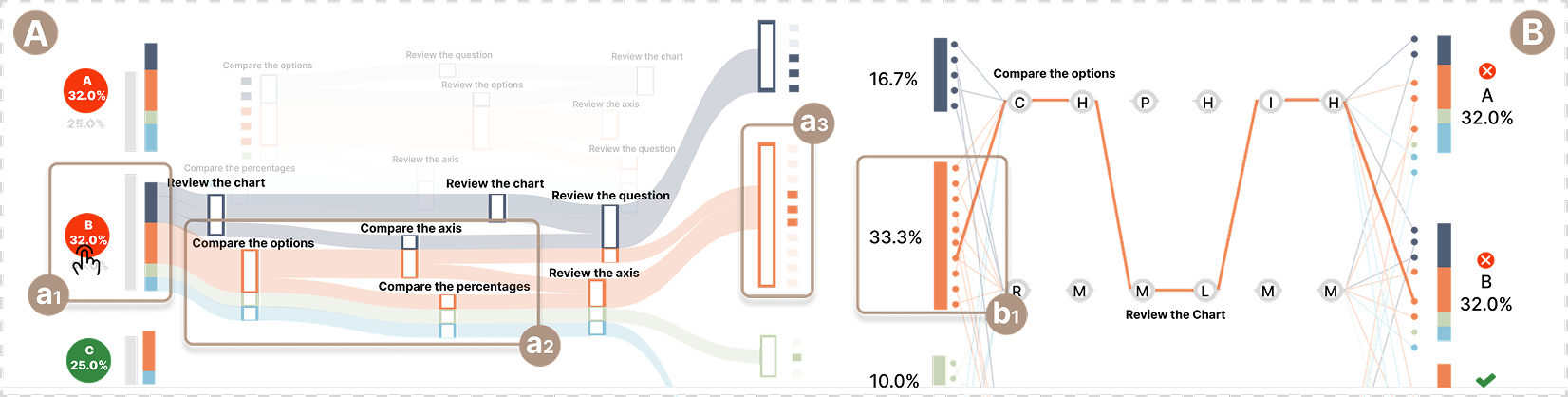}
    \caption{(A) A Sankey-like layout for visualizing student responses and reasoning processes. The colored dot labeled with the option name and number indicates the percentage of students who chose a specific (often incorrect) option. By clicking a particular answer choice (a1), instructors can highlight the relevant subset of students and hide others to reduce visual clutter, thus offering deeper insight into why learners selected certain options and how the question might be improved. A stacked bar chart (a1) shows the composition of each student group, while the Sankey diagram merges students with identical strategies into a single block and uses horizontal spacing to encode average ``thinking time'', approximated by the MLLM's reasoning token length (a2). Each strategy path ultimately reconnects with the student's original group (a3). (B) An alternative node-link design that visualizes reasoning trajectories for different groups. Experts discarded this approach because it does not easily facilitate comparisons of problem-solving processes across students.}
    \label{fig:simulationMain}
\end{figure*}

% \subsubsection{Sample View \& Feature View}
% \label{subsubsec:sample_feature_views}

% Teachers begin by drafting or importing a question in \textbf{Sample View} (\autoref{fig:teaser}-A), where they can either roughly describe the question key points, or paste an existing sample question and attach a reference chart. This initial flexible setup enable teachers to design their question under various situations, and also ensures generating questions aligned with real classroom needs \textbf{(D1)}. 

% Once the question is drafted, teachers move to \textbf{Feature View}  (\autoref{fig:teaser}-B) to fine-tune specific aspects of the MCQ. The system presents four core feature categories, \textit{Question Features, Chart Features, Distractor Features and Knowledge Features} according to the previous collected teachers' needs \textbf{(D2)}. Under each feature category, three to four detailed features are displayed, either in slider or drop down box way to enabling teachers to easily adjust the feature value 
% \textbf{(D3)}. Besides each feature button, a \textit{History} zone displayed the feature value choice of the user's historical choice. Let the users easily see the difference of features among different question versions. At the bottom, the \textit{Generate/Update} will trigger each update of both question and student simulation results in the beside \textbf{Simulation View} (\autoref{fig:teaser}-C).

\subsubsection{Sample View \& Feature View}
\label{subsubsec:sample_feature_views}

% Teachers begin by drafting or importing a question in the \textbf{Sample View} (\autoref{fig:teaser}-A), where they can briefly describe key points or paste an existing sample question and attach a reference chart. This flexible setup allows instructors to design questions for various scenarios while ensuring alignment with real classroom needs \textbf{(R1)}.

After selecting the preferred model via the \textit{Model} button, teachers begin drafting or importing a question in the \textbf{Sample View} (\autoref{fig:teaser}-A), where they can either describe key points or paste sample slides and questions. This flexible setup enables instructors to design questions for various scenarios while ensuring alignment with real classroom needs \textbf{(R1)}.

Once the question is drafted, teachers proceed to the \textbf{Feature View} (\autoref{fig:teaser}-B) to fine-tune specific aspects of the MCQ. The system presents four core feature categories—\textit{Question Features, Chart Features, Distractor Features,} and \textit{Knowledge Features}—based on previously collected teacher needs \textbf{(~\autoref{fig:Feature_Details}-A)}. Within each category, specific features are displayed as sliders or drop-down menus, enabling teachers to easily adjust the relevant values \textbf{(R4)}. Next to each feature control, a \textit{History} section shows previously selected values, allowing users to compare different question versions. Finally, a \textit{Generate/Update} button at the bottom triggers the regeneration of both the question and the student simulation results in the adjacent \textbf{Simulation View} (\autoref{fig:teaser}-C).

\subsubsection{Agent Profile \& Setting Views}
\label{subsubsec:agent_views}
% \ToDo{In Agent View, users can also specify their needs for student agent classification. If no requirements, we use our default settings.}

% To utilize MLLM's module to simulate potential students to test the designed question, teachers could utilize the \textbf{Agent Profile View} (\autoref{fig:teaser}-D) and \textbf{Agent Setting View} (\autoref{fig:teaser}-E). Similar to the \textbf{Sample View} and \textbf{Feature View}, teachers could firstly describe the characteristics distribution of potential students in \textbf{Agent Profile View} (\autoref{fig:teaser}-D), and then allows teachers to fine-tune student agents' setting or emphasize setting focus in \textbf{Agent Setting View} (\autoref{fig:teaser}-E). Here we also provide three main agent setting feature categories with each several detailed setting features according to the literatures and collected teacher needs \textbf{D2-b}. By combining these two views, teachers can model a wide range of learning styles and abilities, creating realistic student simulations that uncover potential pitfalls and support targeted question refinement. The simulated students will be automatically classified into groups and displayed in the \textbf{Simulation View} (\autoref{fig:teaser}-C) for the users to review. 

To simulate potential students using MLLM-based models, teachers rely on the \textbf{Agent Profile View} (\autoref{fig:teaser}-D) and the \textbf{Agent Setting View} (\autoref{fig:teaser}-E). \zixin{Similar to the Sample \& Feature View, instructors can either upload existing student information (e.g., a spreadsheet of demographic attributes and learning characteristics) or begin by describing the broad distribution of student characteristics in the \textbf{Agent Profile View} (\autoref{fig:teaser}-D), then refine and update specific agent settings in the \textbf{Agent Setting View} (\autoref{fig:teaser}-E).} Drawing on prior research and collected teacher needs \textbf{(R2)}, we group agent configurations into three main feature categories (\textit{Demographic Settings}, \textit{Learning Traits Settings}, and \textit{Knowledge Points Settings}), each containing multiple adjustable parameters (\autoref{fig:Feature_Details}-B). By combining these two views, teachers can model diverse learning styles and abilities.  The resulting student groups are automatically classified and displayed in the \textbf{Simulation View} (\autoref{fig:teaser}-C), where instructors can further review each group's performance and feedback.

\begin{figure}[!t]
    \centering
    \includegraphics[width=\linewidth]{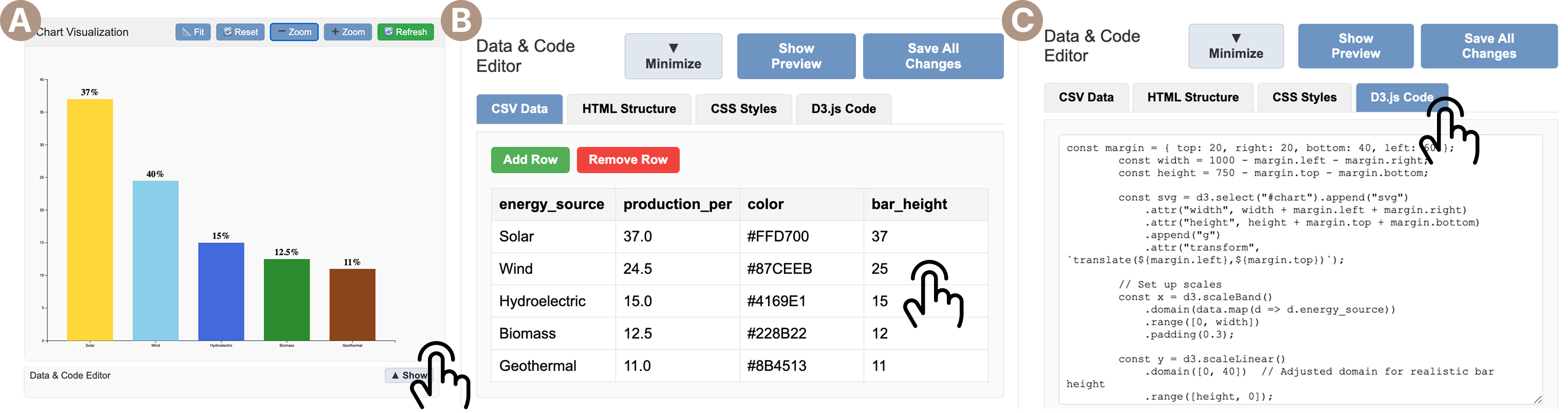}
    \caption{Users can expand the chart editor panel in (A) to review the original CSV data (B) and D3.js/HTML code (C), and edit them directly.
    % By comparing these bars, instructors can quickly assess how changes in the question design or features influence perceived clarity and difficulty across different student groups.
    }
    \label{fig:chartEdit}
\end{figure}

\subsubsection{Simulation View}
\label{subsubsec:simulation_view}

The \textbf{Simulation View} (\autoref{fig:teaser}-C) serves as the central workspace for analyzing both the \emph{Generated Question} and the \emph{Simulation Results}. At the top, the system clusters simulated students into color-coded \emph{Student Groups} (\autoref{fig:teaser}-c1) based on user-defined agent settings. By default, four groups are generated via k-means clustering. Teachers may specify a different number or clustering criteria (\autoref{fig:teaser}-c1). Specifically, each group's panel (\autoref{fig:groupCard}-A) displays essential summaries, such as cluster size, average major or background, and two radar charts illustrating average cognitive and knowledge-point levels.

In the lower half of the \textbf{Simulation View}, the system aligns each component of the \textit{Generated Question} (\autoref{fig:teaser}-c2)—including the stem, chart, answer options, and explanations of relevant knowledge points—on the left, with corresponding student feedback and responses on the right. This side-by-side layout enables instructors to clearly see how settings in the \textbf{Feature View} (~\autoref{fig:teaser}-B) influence each part of the question updates, while also allowing easy comparison with previous versions, indicated by grey bars \textbf{(R4)}.

% This side-by-side arrangement allows instructors to easily track how the feature settings in the \textbf{Feature View} (~\autoref{fig:teaser}-B) influence each part of the question and to compare the results with previous question versions (indicated by grey bars) \textbf{(R4)}.

The simulated students' responses are summarized in the \textit{Simulation Results} zone. At a glance (\autoref{fig:teaser}-c4, \autoref{fig:groupCard}-B), this area displays two rows of stacked bar charts: the top row shows how each student group perceives the question stem, while the bottom row reflects their evaluation of the chart settings, providing a concise summary of student feedback to support quick refinement \textbf{(R3)}. Further down, the Sankey-like diagram (\autoref{fig:teaser}-c5, \autoref{fig:simulationMain}-A) offers deeper insights into students' answer choices and thought processes \textbf{(R3)}. Instructors can click on a specific answer choice (\autoref{fig:simulationMain}-a1) to highlight only the students who selected that option, thereby reducing clutter and revealing shared misconceptions or design flaws. As an example, $32\%$ of students selected the incorrect option B, primarily from the dark-blue and orange student groups (\autoref{fig:simulationMain}-a1). The Sankey-like visualization besides traces each student's reasoning, grouping those who share the same step-by-step strategy into unified blocks, while horizontal spacing encodes the average ``thinking time'' inferred from each agent MLLM's reasoning token lengths (\autoref{fig:simulationMain}-a2). For instance, three students from the orange group who selected option B began by ``Comparing the options''. Two of them then proceeded to ``Compare the axis'', while the third student directly moved on to ``Compare the percentages'' and spent a longer time reasoning through the step. To address potential issues with step length and visual clutter, the backend algorithm incorporates additional control over step length during reasoning analysis. Meanwhile, users can click the ``Full View'' button above the Sankey-like diagram to view an expanded version in a pop-up window. After updating the question feature, by directly editing \autoref{fig:teaser}-c2 or entering natural-language instructions in \autoref{fig:teaser}-c3, teachers can re-run the simulation to observe how these changes affect student responses. A summary chart at the bottom (\autoref{fig:teaser}-c6) compares the current version's correctness rates with previous iterations, enabling quick assessments of how each refinement influences question complexity and clarity \textbf{(R5)}. By aligning question content on the left with student feedback and reasoning on the right, the \textbf{Simulation View} makes it easy to spot ambiguous wording, confusing visuals, or overly challenging knowledge points. Teachers can further revise the question directly or interact with MLLMs in \autoref{fig:teaser}-(c2 and c3). They can also expand the chart editor panel in \autoref{fig:chartEdit}-(A–C) to edit the raw chart code or CSV data if needed. In addition, teachers can adjust features in \autoref{fig:teaser}-B or update student profiles in \autoref{fig:teaser}-(D and E). After making any updates, they can re-run the simulation until the MCQ aligns with instructional goals and addresses the needs of diverse learner populations.

\begin{figure*}[t]
    \centering
    \includegraphics[width=\textwidth]{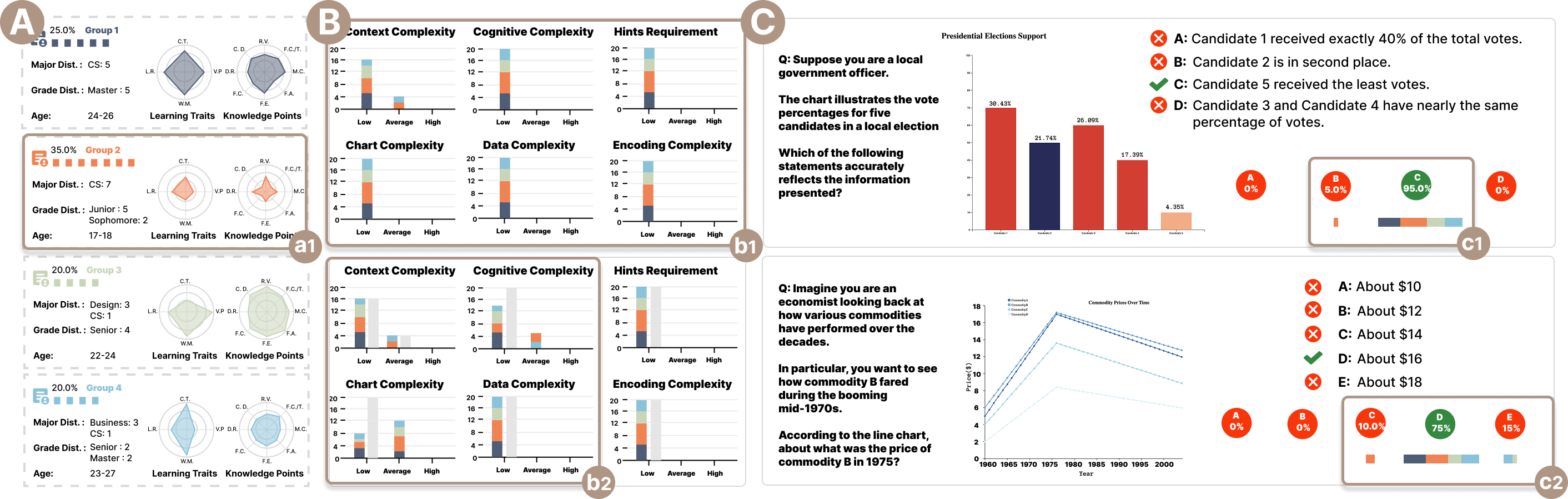}
    \caption{
    (A) Student simulation and clustering results after adjustments by experts E1 \& E5. Two clusters are predominantly CS students with varying ability levels, while the other two clusters comprise Design or Business students, each group specializing in either \textit{Visual Processing} or \textit{Knowledge Points}. (B) Simulated student feedback for the generated questions before and after adjustments. In the upper half, nearly all students rate the original question (top of (C)) as easy across six perspectives. In contrast, after experts revised the chart type and added more distractors (lower half), more simulated students perceive the updated question (bottom of (C)) as more complex. (C) The original (top) and revised (bottom) versions of the expert-generated question.
}
    \label{fig:case1}
\end{figure*}

\textbf{Design alternative.} During the design process, we explored different visualization approaches in consultation with our experts. For example, one idea involved a node-link layout that depicts reasoning trajectories across groups (\autoref{fig:simulationMain}-B). Vertically aligned nodes represent the set of action choices students make at each step, and hovering over a dot (\autoref{fig:simulationMain}-b1) highlights the corresponding problem-solving path. Although this layout more directly reveals individual reasoning, experts ultimately preferred the Sankey-like design because it emphasizes aggregated student performance and facilitates comparisons of problem-solving processes across groups.

\section{Evaluation}
\label{sec:evaluation}

We evaluate {\systemName} via:
(1) a quantitative evaluation of the simulation and question generation modules;
(2) case studies demonstrating system functionalities;
(3) a real-world classroom experiment on designing high-quality visualization literacy MCQs for students; (4) \zixin{a large-scale online study measuring learning gains and pedagogical effectiveness;} and (5) interviews with six domain experts; 

% an online user study gathering feedback from general-public learners on the designed questions.

% In this section, we evaluate {\systemName} through
% (1) a \textit{Quantitative Evaluation} of both the effectiveness and efficiency of the student simulation and question generation modules;
% (2) \textit{Case Studies} illustrating how instructors can leverage the system's core functionalities;
% (3) \textit{Expert Interviews} with six domain experts \textbf{(E1-E6)}; and
% (4) a \textit{Real-World Classroom Experiment} demonstrating how {\systemName} supports instructors in designing high-quality visualization literacy MCQs for diverse student needs.

\subsection{\zixin{Evaluating Student Simulation \& Question Generation}}
\label{subsubsec:agent_evaluation}

\zixin{A primary goal of this evaluation is to assess whether simulated students exhibit reasoning behaviors that are consistent with their intended learner profiles. As prior literature on synthetic users emphasizes, evaluating such agents requires examining adherence to behavioral traits and reasoning processes, rather than task success alone~\cite{hamalainen2023evaluating}. Consequently, general-purpose LLM benchmarks (e.g., MMLU), which focus on overall intelligence~\cite{chang2024survey}, are not suitable for evaluating persona fidelity in student simulation. Our evaluation therefore focuses on two goals: (1) measuring persona fidelity using alignment-based metrics, and (2) informing model selection under interactive design constraints.}

\zixin{\textbf{Alignment-based evaluation of simulation fidelity.}
Following the role-based LLM evaluation framework~\cite{wang2024rolellm}, we operationalized persona fidelity through three complementary alignment dimensions:
\textbf{Cognitive Alignment}, which captures whether the generated reasoning reflects predefined cognitive strengths and weaknesses;
\textbf{Reasoning Steps Alignment}, which assesses whether the reasoning follows expected problem-solving strategies implied by the learner profile (e.g., visual-first vs.\ logic-first approaches);
and \textbf{Semantic Alignment}, which measures the overall thematic coherence between the learner profile description and the generated reasoning trace using semantic textual similarity (STS)~\cite{reimers2019sentence}.
Formal definitions and implementation details of these alignment metrics are provided in the supplementary.}

\zixin{All alignment scores were normalized to the range $[0,1]$ and combined into an overall consistency score using a weighted sum
(40\% cognitive, 40\% reasoning steps, 20\% semantic), deliberately prioritizing process-level fidelity over surface-level semantic similarity. This weighting reflects our design goal of modeling how students reason through visualization tasks, rather than how fluently they describe themselves.}

\zixin{Prior work does not define absolute thresholds for interpreting alignment scores; instead, such metrics serve as \emph{relative indicators} of persona adherence for model comparison and selection~\cite{wang2024rolellm}. In this context, overall consistency values around $0.6$ indicate that a model consistently incorporates profile constraints across repeated generations, rather than producing role-agnostic or highly unstable reasoning traces. We therefore interpret this level of alignment as sufficient to support \emph{design-time exploration} of common reasoning patterns and potential misconceptions, rather than supporting that the simulation accurately predicts individual learner behavior.}

% \zixin{\textbf{Model selection. } Using identical clustered learner profiles and visualization literacy questions (20 simulated students $\times$ 50 rounds), GPT-4o achieved higher overall alignment consistency than o1 (0.6611 vs.\ 0.6237), while also requiring substantially lower latency per response (35.05s vs.\ 146.43s). Given our goal of enabling reasonably stable persona adherence under interactive constraints, we selected GPT-4o as the default simulation model. Detailed per-metric alignment results and runtime analyses are provided in the supplementary.}

\zixin{\textbf{Confirming model selection.}
To validate our design choice under the proposed alignment-based evaluation, we compared GPT-4o with o1 using identical clustered learner profiles and visualization literacy questions (20 simulated students $\times$ 50 rounds).
GPT-4o achieved higher overall alignment consistency than o1 (0.6611 vs.\ 0.6237), while also requiring substantially lower latency per response (35.05s vs.\ 146.43s).
Given our goal of enabling reasonably stable persona adherence under interactive constraints, this result confirms our use of GPT-4o as the default simulation model.}

\zixin{\textbf{Question generation and revision reliability.}
We additionally examined the operational reliability of the question-generation workflow. 
Across 50 randomized generation and revision scenarios, the initial question-generation stage averaged 12.67\,s per instance with a $76\%$ manual success rate, while the revision stage averaged 9.82\,s with an $82\%$ success rate, where success indicates whether the LLM output correctly followed the specified generation or revision instruction.
These results indicate that the system can support timely generation and iterative revision in an interactive authoring context, providing a practical foundation for instructors to iteratively develop and refine their question design.}

\subsection{Case Study}
\label{subsec:case_study}
% \ToDo{check the expert background in formative study; add figures}
We engaged our six experts (E1–E6) from the formative study to evaluate \textit{{\systemName}}. After a brief overview of the system's background and workflow, the experts used \textit{{\systemName}} to iteratively design visualization literacy MCQs. From these sessions, we derived two case studies illustrating the system's utility across diverse scenarios and student backgrounds.

\subsubsection{Iterative Question Design from a Rough Draft}
\label{subsubsec:scenario_iterative_design}

% \begin{figure*}[t]
%     \centering
%     \includegraphics[width=\textwidth]{figs/case_1.png}
%     \caption{
%     (A) Student simulation and clustering results after adjustments by experts E1 \& E5. Two clusters are predominantly CS students with varying ability levels, while the other two clusters comprise Design or Business students, each group specializing in either \textit{Visual Processing} or \textit{Knowledge Points}. (B) Simulated student feedback for the generated questions before and after adjustments. In the upper half, nearly all students rate the original question (top of (C)) as easy across six perspectives. In contrast, after experts revised the chart type and added more distractors (lower half), more simulated students perceive the updated question (bottom of (C)) as more complex. (C) The original (top) and revised (bottom) versions of the expert-generated question.
% }
%     \label{fig:case1}
% \end{figure*}
Experts E1 and E5 used \textit{\systemName{}} to create MCQs starting from a rough idea, iteratively refining each question through feedback and simulation until it met their standards.
% Experts E1 and E5 utilized \textit{\systemName{}} to design MCQs from an initial rough idea, iteratively refining their question through feedback and simulation until a satisfactory version was achieved.

\textbf{Student Profiles Settings and Adjustments.} The experts began by describing their class in the \textbf{Agent Profile View} (\autoref{fig:teaser}-D):
\textit{``My students mostly have a computer science background, with some design and business students. They are new to data visualization with limited foundational knowledge, and their learning traits are average.''}
The initial simulation revealed four clusters, mostly computer science majors with diverse learning traits and visualization knowledge. To better reflect classroom demographics, the experts refined the \textit{Demographic Settings} in the \textbf{Agent Setting View} (\autoref{fig:teaser}-E), boosting the proportion of design and business students and reducing the number of CS students. After these changes, four balanced clusters emerged: two composed mainly of computer science students with varied abilities, one comprising primarily design students strong in \textit{Visual Processing} and certain \textit{Knowledge Points}, and one containing mostly business students with higher \textit{Critical Thinking} (~\autoref{fig:case1}-A).
% The experts began by providing a general description of the class in the \textbf{Agent Profile View} (~\autoref{fig:teaser}-D): \textit{``My students mostly have a computer science background, with some design and business students. Generally, they are new to data visualization with limited foundational knowledge, and their learning traits are average.''} The simulation results initially indicated four student clusters, predominantly consisting of computer science students with diverse learning traits and varying visualization literacy levels. To better reflect classroom demographics, experts refined the \textit{Demographic Settings} within the \textbf{Agent Setting View} (~\autoref{fig:teaser}-E), slightly increasing the representation of design and business students while reducing the proportion of computer science students. After these adjustments, the simulated student groups were well-balanced across four clusters: two groups primarily contained computer science students with varied abilities, one group consisted largely of design students with strengths in \textit{Visual Processing} and specific \textit{Knowledge Points}, and the remaining group included mostly business students with stronger \textit{Critical Thinking} traits (~\autoref{fig:case1}-A).

\textbf{Rough Question Description and Iterative Adjustments.} With student profiles set, the experts moved to the \textbf{Sample View} (\autoref{fig:teaser}-A) to outline their question requirements. They started with a broad description:
\textit{``I need a bar chart question to retrieve and compare values among different groups. It should not be overly straightforward. Include an interesting contextual background.''}
They kept the \textbf{Feature View} (\autoref{fig:teaser}-B) mostly at default, modifying only the \textit{Chart Type}.

% With student profiles confirmed, the experts proceeded to the \textbf{Sample View} (~\autoref{fig:teaser}-A) to articulate their question requirements. Since they initially had only a vague idea, they provided a general description: \textit{``I need a bar chart question asking students to retrieve and compare values across different groups. It should not be overly straightforward and might include an interesting contextual background.''} The experts maintained default settings in the \textbf{Feature View} (~\autoref{fig:teaser}-B), adjusting only the \textit{Chart Type} selection.

After the initial question generation and student response simulation, the experts reviewed the output (upper half of \autoref{fig:case1}-B,C). They confirmed the question's correctness and overall quality; simulated student feedback (\autoref{fig:case1}-b1) indicated that most aspects, such as chart interpretation, were easy, with moderate complexity noted in context—matching the experts' expectations. However, the initial accuracy rate was too high at $95\%$ (\autoref{fig:case1}-c1), suggesting insufficient difficulty. Only one student from group 2, a junior with notably low traits, answered incorrectly (\autoref{fig:case1}-a1). To raise the challenge level, the experts revisited the \textbf{Feature View} (\autoref{fig:teaser}-B), increasing both \textit{Number of Distractors} and \textit{Plausibility}, and switching the \textit{Chart Type} to a line chart for additional complexity. Upon regeneration and simulation, the updated question offered four plausible distractors, reducing the accuracy to a more desirable $75\%$ (\autoref{fig:case1}-c2). The newly introduced distractors involved visually similar line charts, requiring more careful inspection according to the updated student feedback (~\autoref{fig:case1}-b2). Satisfied with the new difficulty level, the experts marked the question as ``Checked'', finalizing it for classroom use.

\subsubsection{Designing Misleading MCQs for Educational Purposes}
\label{subsubsec:scenario_personalized_learning}

\begin{figure}[!b]
    \centering
    \includegraphics[width=\linewidth]{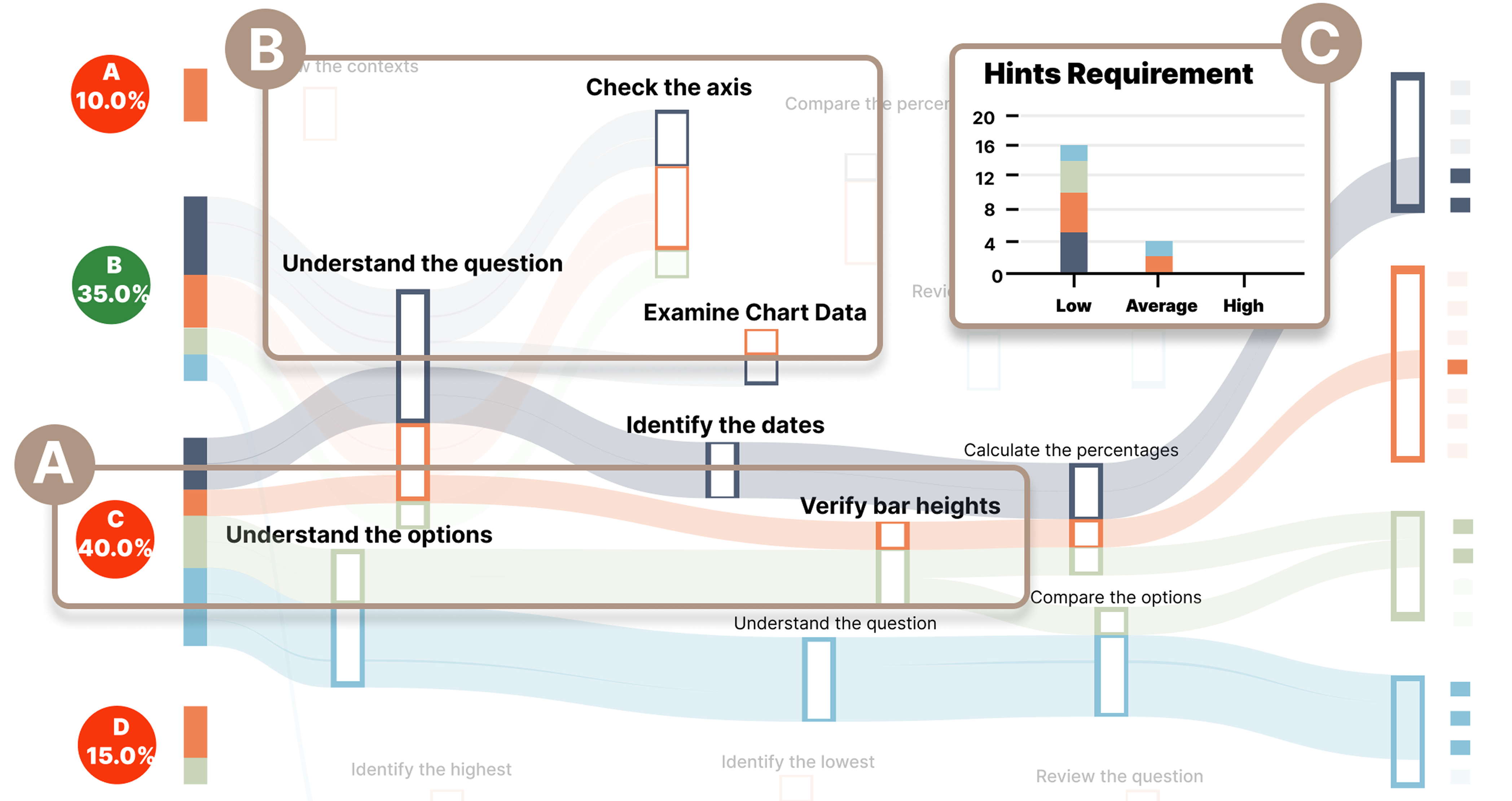}
    \caption{Simulation results for the first version of the misleading bar chart question in Section 6.2.2, where simulated students achieved a $35\%$ correct rate. (A) Students who fell into the ``trap'' and chose option C tended to check each bar's height directly after understanding the question and options. (B) Students who answered correctly typically examined the chart axis or data. (C) Most simulated students overlooked the truncated axis and considered the question simple; they generally rated the hint requirement as low.
}
    \label{fig:case2}
\end{figure}

% Experts E2 and E3, currently teaching an undergraduate-level data visualization course, leveraged {\systemName} to generate and iteratively refine questions derived from their existing lecture materials, tailored specifically for their students.

Experts E2 and E3, instructors of an undergraduate data visualization course, used \textit{\systemName} to iteratively design questions based on their lecture materials, tailoring them to the specific needs of their students.

\textbf{Slide Upload and Requirement Specification.} Seeking to emphasize the misleader \textit{Inappropriate Scale Range}—where a non-zero y-axis skews the perceived proportions in a bar chart—they uploaded a related slide screenshot (provide in supplementary materials) and provided clear instructions:
\textit{``I plan to create a misleading bar chart question that does not start at zero, thus skewing the perceived proportion between two items.''}
After confirming that the simulated student profile clusters matched their class composition, the experts proceeded to simulate student responses for their initial draft.

% The experts aimed to design a question emphasizing the common visualization misleader \textit{Inappropriate Scale Range}, where a y-axis not starting from zero typically causes misinterpretations regarding proportions between bar heights. To initiate this process, they uploaded a relevant slide screenshot (see the supplementary) and provided a clear instructional prompt: \textit{“I plan to create a misleading bar chart question that does not start at zero, thus skewing the perceived proportion between two items.”} After verifying that the generated student profile clusters matched their actual class composition, the experts proceeded to simulate student responses to the initial question draft.

\textbf{Identification of Common Error Reasoning.} The generated question (~\autoref{fig:teaser}-c2) aligned well with the experts' expectations and contained no clear errors. However, while the simulated students found it neither overly complex nor excessively tricky, the overall simulated accuracy rate was only $35\%$ (\autoref{fig:case2}). To understand why, the experts examined detailed reasoning traces (\autoref{fig:case2}-A). They saw that students who answered incorrectly typically proceeded directly to \textit{Verify bar heights} right after \textit{Understand the question} or \textit{Understand the options}, neglecting the axis scale. In contrast, those who answered correctly either performed \textit{Examine Chart Data} or explicitly \textit{Checked the Chart Axis}, revealing distinctly different reasoning strategies (\autoref{fig:case2}-B). Although few students indicated the need for hints (\autoref{fig:case2}-C), the experts decided to include a subtle hint to draw attention to the axis scale, reinforcing this crucial visualization literacy concept rather than simply tricking students.

% The generated question (see Figure~\ref{fig:teaser}-c2) matched the experts' initial expectations, presenting no errors. Simulated student feedback indicated that the question was neither excessively complex nor deliberately tricky. However, the overall simulated accuracy rate ($35\%$) was significantly lower than expected (~\ref{fig:case2}). To investigate further, the experts reviewed detailed reasoning traces from the simulation (~\ref{fig:case2}-A). They discovered that students who answered incorrectly typically proceeded directly to \textit{Verify bar heights} directly after \textit{Understand the options} or \textit{Understand the question}, neglecting the misleading axis scale. Conversely, students who answered correctly either performed \textit{Examine Chart Data} or explicitly \textit{Checked the Chart Axis}, underscoring a clear difference in reasoning strategies (~\ref{fig:case2}-B). Although few students explicitly indicated needing a hint (~\ref{fig:case2}-C), the experts decided that including a subtle hint to guide attention towards the axis scale was necessary—not to intentionally trick students but to reinforce their understanding of this crucial visualization literacy concept.

\textbf{Question Revision and Resimulation.} Returning to the question revision panel (\autoref{fig:teaser}-c3), the experts added a textual instruction:
\textit{``Add a subtle hint to the question stem.''}
After regenerating the question with this minor change (\autoref{fig:teaser}-c7), a resimulation showed a clear improvement: more than $65\%$ of students now correctly recognized the misleading element by inspecting the axis scale before comparing bar heights (~\autoref{fig:teaser}-c5). Notably, simulated student feedback (~\autoref{fig:teaser}-c4) explicitly mentioned the increased recognition and utility of this hint. Both the original and revised questions were saved, reinforcing the value of guiding hints in this challenging question.

% Consequently, the experts returned to the question revision panel (~\ref{fig:teaser}-c3) and input the textual instruction: \textit{“Add a subtle hint to the question stem.”} After regenerating the question with only this minor modification (Figure~\ref{fig:teaser}-c7), a resimulation indicated a noticeable improvement, with more than $65\%$ of students now correctly identifying the misleading element by carefully examining the axis scale before comparing bar heights (Figure~\ref{fig:teaser}-c5). Interestingly, simulated student feedback (Figure~\ref{fig:teaser}-c4) also explicitly highlighted increased recognition of the necessity and usefulness of the provided hint. This reinforced the experts' decision to include guiding hints in challenging visualization literacy MCQs. Ultimately, both the original and revised question versions were saved for future instructional use.

\subsection{Real-World Classroom Experiment}
\label{subsec:class_experiment}
After the expert interviews, we conducted a real-world classroom experiment (IRB-approved) to evaluate \textit{\systemName{}} by assessing the quality of MCQs designed with the system from students' perspectives and examining how well its simulated student performance aligns with real student outcomes. Collaborating with E3, we designed an in-class quiz for a lecture on \textit{Effective Visual Design for Data Storytelling}.
% After the expert interviews, we conducted a real-world classroom experiment to evaluate \textit{\systemName{}} in an undergraduate-level data visualization course taught by expert E3. Collaborating with E3, we designed an in-class quiz for a lecture on \textit{Effective Visual Choices for Storytelling}.

% Following the expert interviews, we conducted a real-world classroom experiment to evaluate \textit{\systemName{}} within an undergraduate-level data visualization course taught by expert E3. Specifically, we collaborated with E3 to design an in-class quiz for a lecture titled \textit{Effective Visual Choices for Storytelling}.

\begin{figure}[!t]
    \centering
    \includegraphics[width=\linewidth]{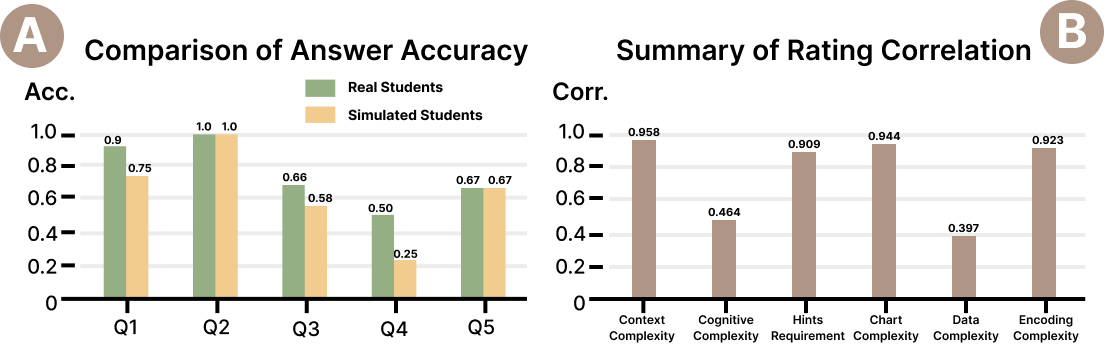}
    \caption{
    (A) Comparison of real students' answer accuracy with that of the simulated student agents (designed by expert E3) for each question. See the supplementary materials for details on the questions. (B) Pearson correlation coefficients summarizing the alignment between simulated agents' and real students' ratings on five feedback metrics, aggregated across all questions.
}
    \label{fig:statistics}
\end{figure}

\textbf{Quiz Design and Setup.} E3 sought to design MCQs that assess four critical knowledge points from the lecture: (1) effective color schemes for heatmaps, (2) scatterplots for visualizing variable relationships, (3) reduce visual clutter in line charts, and (4) potential misinterpretation with bar charts when the y-axis does not start at zero. Initially, E3 included two previously designed misleading bar chart questions—one with explicit hints and one without—to explore whether hints effectively guided students to identify misleading elements. Next, E3 used \textit{\systemName{}} to upload relevant lecture slides and initial question descriptions, iteratively refining the questions based on simulated student feedback. This process balanced overall difficulty and yielded a final quiz of five questions. The students took this quiz during the following lecture, and the participation was voluntary, anonymous and IRB approved. A total of 12 students agreed to share data for research purposes. They rated each question on six dimensions aligned with our simulation metrics (\autoref{subsec:student_simulation}-Simulating Student Responses), giving comments to the quiz question design, providing insights into both question quality and how well the simulated agents reflected the real student experiences.

\textbf{Student Feedback and Agent Comparison.}
Overall, student feedback was very positive. Many noted strong alignment with the class materials, for instance:
\textit{``They aligned with the class materials. I can see some of the heuristics and biases used in the graph design to manipulate and mislead data.''}
Others praised the visual clarity, calling it ``visually pleasing'' and describing the graphs as ``interesting and nicely displayed''. We compared each question's average accuracy between real students and the simulated agents designed by E3, and also computed Pearson correlations between real student ratings and simulated student feedback (\autoref{fig:statistics}). The results indicate that the simulated agents generally predicted student performance accurately, although they tended to slightly underestimate real students' success (\autoref{fig:statistics}-A). As for the agents' modeling of student ratings, performance varied across metrics (\autoref{fig:statistics}-B): the agents were more accurate in capturing context and visual comprehension, while relatively weaker in modeling data complexity or overall cognitive challenges.

In summary, our experiment demonstrates the effectiveness and reliability of our simulated student module in capturing genuine student perceptions and providing valuable guidance for MCQ refinement.

\subsection{\zixin{Large-Scale Online Study on Learning Outcomes and Simulation Validity}}
\label{subsec:online_study}

\zixin{We conducted a large-scale online study to examine the educational impact of \systemName{} and the validity of its student simulation in supporting instructional design. 
The study addressed two research questions: 
(1) whether MCQs designed with the support of \systemName{} can function as effective instructional materials and lead to measurable learning outcomes in visualization literacy; and 
(2) whether the LLM-simulated reasoning strategies used during question design can credibly approximate common learner thinking patterns.}

\zixin{\textbf{Study Design Overview.} The study followed a two-stage design that mirrors a realistic instructional workflow: learner profiling, question design, instruction with feedback, and post-instruction assessment. Across all stages, learning objectives were controlled by focusing on a fixed set of five visualization misleaders, which served as the core knowledge construct.}

\zixin{\textbf{Stage~1: Learner Profiling, Baseline Assessment, and Group Assignment.}
In Stage~1, $100$ participants were recruited via Prolific, of whom $99$ provided valid responses. Participants reported background information and completed self-report items assessing cognitive and visualization-related abilities, aligned with the learner attributes used in \systemName{}'s student simulation. Participants also completed a five-item visualization literacy pre-test (score range: $0$--$5$), with instructor-selected VLAT questions aligned with the visualization misleaders targeted in the tutorial.}

\zixin{Based on pre-test performance and demographic characteristics, participants were divided into two groups with comparable baseline scores and similar demographic distributions (Group~1: $N=50$, Mean $=2.10$, SD $=1.15$; Group~2: $N=49$, Mean $=2.12$, SD $=1.17$). 
To reduce memorization effects, pre-test items instantiated the target misleaders using chart types that differed from those used in later stages.}
% \ToDo{add a figure about pre-demographic abilities score distribution}

\zixin{\textbf{Teacher-Guided Question Design with \systemName{}.} Instructors who had previously taught the visualization literacy course were invited to design instructional MCQs using \systemName{}. They were provided with the learner profiles of Group~1 and designed ten MCQs targeting the five misleaders (two questions per misleader). During this process, \systemName{} generated simulated reasoning traces for all 50 learners per question, recorded instructors' edit logs, and clustered simulated reasoning to extract the five most frequent reasoning patterns per question.}

% \zixin{\textbf{Stage~2: Concept Instruction and MCQ-Based Learning Activities.} Stage~2 was conducted three days after Stage~1. Participants were given a one-week window to complete Stage~2. All participants first reviewed a tutorial introducing the visualization literacy concepts associated with the five misleaders. Participants then completed a learning activity consisting of ten MCQs, which functioned as learning materials: after each response, participants were shown the correct answer and an explanation.}

% \zixin{Participants in Group~1 completed the ten MCQs designed with \systemName{}, while Group~2 completed ten VLAT-aligned MCQs covering the same knowledge points. The underlying misleaders were consistent across pre-test, instructional MCQs, and post-test, while chart types were varied to promote conceptual transfer. After completing the learning activity, participants completed a five-item visualization literacy post-test. Of the $99$ Stage~1 participants, $80$ completed Stage~2 with consent and tutorial completion, resulting in $39$ valid responses in Group~1 and $41$ in Group~2.}

\zixin{\textbf{Stage~2: Concept Instruction, MCQ-Based Learning, and Reasoning Alignment Tasks.} 
Stage~2 was conducted three days after Stage~1, and participants were given a one-week window to complete all tasks. All participants first reviewed a tutorial introducing the visualization literacy concepts associated with the five misleaders, ensuring a shared instructional baseline.}

\zixin{Participants then completed a learning activity consisting of ten MCQs, which functioned as instructional materials rather than solely as assessment items. After responding to each question, participants were shown the correct answer along with an explanation to support conceptual understanding.} 

\zixin{Participants in Group~1 completed the ten MCQs designed using \systemName{}, while participants in Group~2 completed ten VLAT MCQs covering the same knowledge points. 
The underlying misleaders were consistent across the pre-test, instructional MCQs, and post-test, while chart types and visual encodings were varied to promote conceptual transfer. For participants in Group~1, an additional reasoning alignment task was included after answering each MCQ. Specifically, participants were presented with the five most frequent reasoning strategies derived from clustering the LLM-simulated student reasoning traces for that question. Participants were asked to select the strategy that most closely matched their own thinking process. If none of the strategies adequately reflected their reasoning, participants could select an ``Other'' option and describe their reasoning in free text. After completing the learning activity, all participants completed a five-item post-test, consisting of VLAT questions selected to align with the same knowledge points assessed in the pre-test.}

\zixin{Of the $99$ participants who completed Stage 1, $80$ proceeded to Stage 2 with valid consent and tutorial completion, yielding $39$ valid responses in Group 1 (11 incomplete submissions) and $41$ in Group 2 (8 incomplete submissions).
}

% \zixin{\subsubsection{Learning Outcomes and Learning Gains} Participants demonstrated improved visualization literacy following instruction. 
% Post-test performance was comparable across conditions (Group~1: Mean $=2.95$, SD $=1.30$, $N=39$; Group~2: Mean $=2.88$, SD $=1.31$, $N=41$). Using matched pre--post data ($N=79$), both groups exhibited significant learning gains. 
% Group~1 showed a mean gain of $0.74$ (SD $=1.67$, $N=38$; paired $t$-test, $p=0.010$), while Group~2 showed a mean gain of $0.63$ (SD $=1.37$, $N=41$; paired $t$-test, $p=0.005$). Across all participants, the average learning gain was $0.68$ (SD $=1.52$), with a 95\% confidence interval of $[0.34,\,1.02]$.}

% To compare instructional conditions while accounting for baseline differences, we conducted an ANCOVA with post-test score as the dependent variable and pre-test score as a covariate. 
% The analysis revealed no significant effect of condition (group coefficient $=-0.04$, $p=0.887$), indicating that system-supported instructional MCQs facilitated learning outcomes comparable to VLAT-aligned materials.

\subsubsection{\zixin{Learning Outcomes and Learning Gains}}

\begin{figure}[h]
    \centering
    \includegraphics[width=\linewidth]{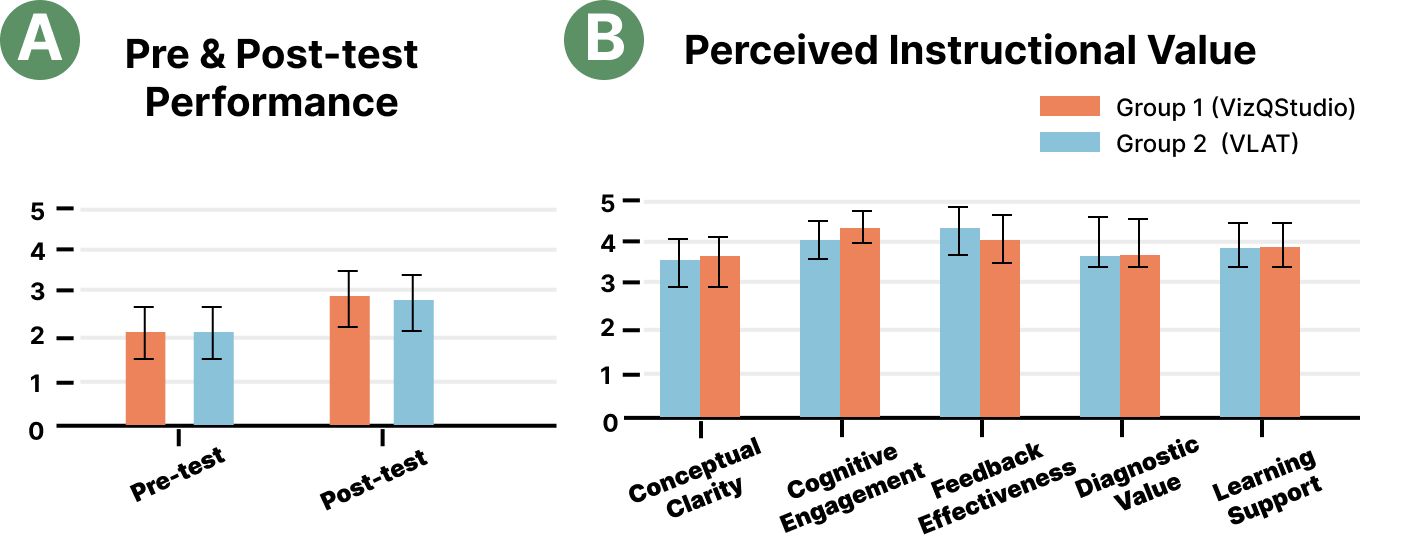}
    \caption{(A) Pre-test and post-test performance of participants in the two instructional conditions. (B) Participants’ perceived instructional value of the MCQs across multiple dimensions for \systemName{}-designed questions (Group~1) and benchmark VLAT questions (Group~2).
}
    \label{fig:onlineStudy}
\end{figure}

\zixin{Participants demonstrated improved visualization literacy following instruction in both conditions. 
Post-test performance was comparable across conditions (Group~1: Mean $=2.95$, SD $=1.30$, $N=39$; Group~2: Mean $=2.88$, SD $=1.31$, $N=41$) (\autoref{fig:onlineStudy}-A), suggesting that learners were able to demonstrate similar levels of visualization literacy after engaging with either set of instructional MCQs.}

\zixin{Using matched pre--post data ($N=79$), both groups exhibited statistically significant learning gains relative to their baseline performance. Group~1 showed a mean gain of $0.74$ (SD $=1.67$, $N=38$; paired $t$-test, $p=0.010$), while Group~2 showed a mean gain of $0.63$ (SD $=1.37$, $N=41$; paired $t$-test, $p=0.005$). 
Across all participants, the average learning gain was $0.68$ (SD $=1.52$), with a 95\% confidence interval of $[0.34,\,1.02]$, indicating a consistent improvement in visualization literacy following the instructional activities.}

\zixin{To compare instructional conditions while accounting for baseline differences, we conducted an ANOVA with post-test score as the dependent variable and pre-test score as a covariate. 
The analysis revealed no significant effect of condition (group coefficient $=-0.04$, $p=0.887$). Rather than aiming to outperform standardized instruments such as VLAT, this analysis evaluates whether MCQs authored with \systemName{} can achieve comparable learning outcomes under substantially more flexible design conditions. The absence of a significant difference indicates that, despite relying on instructor-driven generation and simulation rather than a fixed, curated item bank, \systemName{}-designed questions do not compromise learning effectiveness. This result points to the \systemName{}'s potential role at design time: supporting scalable, low-cost, and adaptable question authoring in instructional contexts where standardized tests are impractical due to curriculum mismatch, limited coverage, or the need for frequent iteration.}

\subsubsection{\zixin{Validity of LLM-Simulated Reasoning Strategies}}

\zixin{A central goal of \systemName{} is to support instructional design by simulating plausible student reasoning patterns. 
Accordingly, our evaluation focuses on whether LLM-simulated reasoning strategies can adequately represent common student thinking patterns observed during problem solving.}

\zixin{\textbf{Coverage of Student Reasoning Patterns.}
Across $39$ participants and $10$ questions in Group~1 (i.e., $390$ reasoning alignment instances), $94.62\%$ of responses selected one of the five representative reasoning strategies derived from clustering the simulated reasoning traces. 
At the question level, coverage exceeded $92\%$ for all items, indicating that the majority of participants were able to identify a simulated reasoning strategy that closely matched their own thinking process.}

\zixin{\textbf{Perceived Alignment with Simulated Reasoning Strategies.}
In addition to selecting a matching strategy, participants rated how well the provided strategies aligned with their own reasoning process on a five-point Likert scale. 
Across all questions and participants in Group~1, participants reported a generally high perceived alignment between the simulated reasoning strategies and their own thinking (Mean $=3.62$, SD $=1.14$, $N=39$), indicating that the LLM-simulated strategies were often perceived as capturing key aspects of students' reasoning processes. This quantitative assessment complements the coverage results by capturing the degree of perceived similarity between simulated and self-reported student reasoning.}

\zixin{\textbf{Analysis of Misalignment Cases.}
In cases where participants selected the ``Other'' option or reported lower alignment, free-form responses revealed that misalignment typically stemmed from heuristic estimation (e.g., approximate visual comparison without explicit stepwise reasoning), individual visual accessibility issues, or idiosyncratic reasoning shortcuts. 
For example, one participant noted that they ``roughly compared the bar lengths visually and chose the closest option,'' without explicitly following the multi-step comparison described in the simulated strategies. 
These responses suggest that mismatches often reflected differences in the level of reasoning granularity or articulation, rather than fundamentally distinct problem-solving approaches.}

% Another participant described first filtering visually salient elements before reasoning about values, a step that was not explicitly represented in the provided strategies. 

\zixin{\textbf{Relationship Between Simulation Alignment and Learning Outcomes.}
To examine whether alignment with simulated reasoning strategies relates to learning outcomes, we conducted exploratory analyses within Group~1. 
Neither participants' average alignment ratings nor the frequency of ``Other'' selections showed a significant Pearson's correlation with post-test performance ($|r| < 0.25$, $p > 0.20$)~\cite{benesty2009pearson}. This result suggests that alignment with simulated reasoning strategies is not strongly associated with individual learning outcomes, and that the primary role of the simulation lies in supporting instructional design rather than predicting learner performance.}

% \ToDo{Move to discussion about the simulation limitation}

% \subsubsection{\zixin{Perceived Instructional}}

% \zixin{Participants reported positive perceptions of the MCQs as learning materials following the tutorial in both conditions. 
% For MCQs designed with \systemName{} (Group~1), participants rated the questions as clear (Conceptual Clarity: Mean $=3.54$, SD $=1.19$), cognitively engaging (Cognitive Engagement: Mean $=3.97$, SD $=0.99$), and effective in supporting understanding through answers and explanations (Feedback Effectiveness: Mean $=4.15$, SD $=1.04$), indicating strong perceived instructional value when the questions were used as guided learning activities.}

% \zixin{To contextualize these perceptions, we compared Group~1 with a VLAT-aligned benchmark condition (Group~2). Across the same dimensions, Group~2 exhibited overall comparable ratings (Conceptual Clarity: Mean $=3.59$, SD $=1.26$; Cognitive Engagement: Mean $=4.17$, SD $=0.77$; Feedback Effectiveness: Mean $=4.05$, SD $=0.97$). Taken together, these results suggest that the two sets of MCQs were perceived as offering comparable levels of clarity, engagement, feedback support, and overall instructional value. This indicates that teachers using \systemName{} can design instructional MCQs that learners perceive as effective learning materials, with educational value comparable to VLAT-aligned questions in a tutorial-based learning context.}

\subsubsection{\zixin{Perceived Instructional Value}}

\zixin{Participants reported positive perceptions of the MCQs as learning materials following the tutorial (\autoref{fig:onlineStudy}-B). 
For the \systemName{}-designed MCQs (Group~1), participants rated the questions as clear (Conceptual Clarity: Mean $=3.54$, SD $=1.19$) and cognitively engaging (Cognitive Challenge / Engagement: Mean $=3.97$, SD $=0.99$), indicating that the teacher-authored questions produced with \systemName{} were accessible while still prompting active reasoning. 
Participants also reported that the provided answers and explanations effectively supported understanding (Feedback Effectiveness: Mean $=4.15$, SD $=1.04$).}

\zixin{Beyond surface-level clarity and engagement, participants perceived the \systemName{}-designed MCQs as instructionally meaningful in diagnosing and supporting learning. 
Specifically, participants reported that the questions helped surface misunderstandings (Diagnostic Value: Mean $=3.64$, SD $=1.20$) and supported preparation for answering similar visualization literacy questions (Learning Support / Preparedness: Mean $=3.79$, SD $=1.08$). 
These perceptions suggest that the MCQs functioned not only as practice items but also as reflective learning prompts that encouraged learners to examine their reasoning.}

\zixin{To contextualize these perceptions, we compared Group~1 with a VLAT benchmark condition (Group~2), which exhibited overall comparable ratings across the same instructional-value dimensions (\autoref{fig:statistics}-B) (Conceptual Clarity: Mean $=3.59$, SD $=1.26$; Cognitive Engagement: Mean $=4.17$, SD $=0.77$; Diagnostic Value: Mean $=3.68$, SD $=1.11$; Feedback Effectiveness: Mean $=4.05$, SD $=0.97$; Learning Support: Mean $=3.83$, SD $=1.14$). 
Taken together, these results indicate that teachers using \systemName{} can design instructional MCQs that learners perceive as effective learning materials, with overall instructional value comparable to VLAT questions in a tutorial-based learning context.}

\subsubsection{\zixin{Teacher Revision Patterns and Human-in-the-Loop Safeguards}}

\zixin{During teacher-guided question design, \systemName{} recorded detailed revision logs capturing how instructors iteratively refined each MCQ prior to deployment. Across the ten instructional MCQs designed for Stage~2, instructors revised each question multiple times, with an average of $8.3$ revisions per question (SD $=2.4$) and an average authoring time of $9.6$ minutes per question (SD $=3.1$).}

\zixin{Based on revision logs and instructor feedback, approximately half of the revisions focused on structural aspects of the question, such as modifying chart data or visual encodings to ensure correctness and clarity, while the remaining revisions were driven by inspecting simulated student reasoning and adjusting difficulty, distractors, or explanations accordingly. Notably, although chart-related revisions accounted for a substantial portion of revision events, they consumed a smaller share of total authoring time (approximately one-third), facilitated by flexible editing affordances such as direct chart manipulation and natural-language interaction with the LLM.}

\zixin{Together, these results indicate that \systemName{} supports an iterative, expert-driven design process in which model-generated drafts are efficiently corrected and pedagogically refined through simulation-informed exploration. Broader implications and limitations are discussed in Sec.~\ref{sec:discussion}.}

\subsection{Expert Interviews}
\label{subsec:expert_interviews}
Following the case studies, we conducted one-on-one interviews with six experts (E5–E10, including two from the case studies and four newly invited assistant professors specializing in visualization from two universities). Each interview lasted 60–80 minutes with a $\$70$ honorarium. For the newly invited experts, the session included a background introduction, system demonstration, hands-on trial, and feedback focusing on usability, workflow, and simulation reliability.

% Following the case studies, we conducted one-on-one interviews with six experts \revision{(E4-E10, two where the same from the case study and four new university assistant professors with expertise in visualization from two universities)}, each lasting 60-80 minutes with a $\$70$ compensation, to gather professional perspectives on {\systemName}. For the newly invited experts, the study consist of a background introduction, followed up with a in detailed system demo and explanation, then the experts will also try out the system and giving feedback targeting mainly at system design and usability, workflow and realiability.

\textbf{System Workflow.} All experts praised {\systemName}'s workflow for its clarity, flexibility, and functionality. They particularly valued the system’s versatile question-generation methods, such as transforming rough inputs (e.g., lecture slide screenshots) into well-aligned questions targeting specific knowledge points. Experts also highlighted its intuitive editing features for both questions and student agents, support for free-form exploration in question creation, seamless integration with real-time simulation updates through a clear side-by-side interface, and robust version-saving capabilities. These features significantly streamlined the iterative design process.

\textbf{Question Generation and Student Simulation Modules.} All experts highlighted the \textit{Question Generation} and \textit{Student Simulation} modules as effective, stable, and insightful. They reported fast runtimes — approximately 20 seconds for question generation, 10 seconds for each LLM-driven update, and 30 seconds for student simulation with a typical class size of 30–40 students. Experts were particularly impressed by the quality and flexibility of the generated questions. For example, E8 remarked, \textit{``The generated questions are impressive. Even with just a rough idea or a slide, the system accurately captures my intent and produces charts and QAs with clear explanations. The editing panels further ensure the reliability of the questions while giving me fine-grained control.''}

% All experts noted the \textit{Question Generation} and \textit{Student Simulation} modules as effective, stable, and insightful. They reported quick runtimes—around 20 seconds for question generation, about 10 seconds for each LLM-driven question updates, and roughly 30 seconds for student simulation results under a regular 30-40 size. Experts were particularly impressed with the high quality and flexibility of generated questions. For example, E8 observed, \textit{``The generated question quality is very impressive. Even with just a rough idea or a slide, the system accurately captures my intent and produces correct charts and QAs with clear explanations. The editing panels further made the question reliable with my control.''}. 

Experts highly praised the student simulation module for its potential to provide insights into students’ reasoning across different backgrounds, which can particularly help them design and test questions tailored to students’ needs. As E10 remarked, \textit{``I intentionally adjusted two groups of students — one with high general learning traits but low visualization knowledge, and the other with the opposite profile — to observe differences in their answers and reasoning. The results were impressive and could help me design more targeted questions for different groups, anticipate how they might think, and assess whether the question sets effectively support their intended learning outcomes.''}

\textbf{Visual Designs and Interactions.} Experts consistently described the system’s visual design and interactions as clear and intuitive. The stacked bar charts were especially helpful for quickly gauging overall student feedback on the question, while the Sankey-like diagram effectively illustrated each group's distribution of final answers and reasoning pathways. As E9 noted, \textit{``my top priorities are students’ perceptions and overall question difficulty, and the system effectively summarizes this while still allowing me to inspect each student’s reasoning.''} The rich interaction methods also offered great flexibility, enabling experts to design and edit questions and student profiles through natural language input, button and slider interactions, as well as direct editing.

\textbf{Suggestions.} Experts offered several ideas for enhancing {\systemName}. E9 suggested reducing the number of default options in question features and providing full options only on demand to simplify the system. E10 and E7 recommended giving greater prominence to the overall question difficulty chart, as they tend to examine the difficulty across the entire question set first. E8 further proposed strengthening the linkage between the reasoning steps in the Sankey diagram and the learning traits and knowledge points, which could provide deeper insights into different student groups.

\section{\zixin{Discussion}}
\label{sec:discussion}

\zixin{This section synthesizes insights from our design and evaluation, situates our approach within broader educational and learner modeling research, and discusses limitations, ethical considerations, and implications for future work.}

\subsection{\zixin{LLMs in Assessment Design: Multi-level Human Oversight as a Design Necessity.}}
\zixin{While LLMs can effectively generate diverse question drafts, distractors, and plausible reasoning pathways, in practice they are most appropriately positioned as generative and exploratory tools. Correctness, pedagogical alignment, and final validation must remain under instructor control.}

\zixin{A central insight from our study is that effectively leveraging LLMs for assessment design requires multi-level instructor supervision. At a \textbf{higher level}, instructors benefited from interaction paradigms that allow them to communicate desired changes to LLMs in ways that resemble authentic teaching practice—for example, describing issues in natural language, requesting alternative formulations, or indicating (e.g., through sketching, annotation, or highlighting) which parts of a question or visualization require revision. Such interaction supports fluid exploration and refinement, enabling instructors to adjust LLM-generated question framing, distractors, or explanations without directly engaging with technical representations.}

\zixin{At the same time, our observations highlight that high-level interaction alone is insufficient for high-stakes educational materials. Instructors required precise, deterministic control over the final learning artifacts to ensure factual and conceptual correctness, particularly for visualization-based questions where small changes in data values, scales, or encodings can alter the underlying concept being assessed. This necessitates \textbf{lower-level controls} that allow instructors to directly verify and modify the concrete components of assessment materials.}

\zixin{In addition, grounding LLM generation in domain-specific references emerged as an important complement to both high- and low-level oversight. Incorporating instructor-provided examples, prior assessments, or openly available educational resources—together with techniques such as retrieval-augmented generation—helps anchor generation in curricular context and mitigates limitations of generic model knowledge.}

\zixin{Taken together, these insights suggest a broader design implication: in educational settings, effective LLM-supported authoring hinges not on minimizing human involvement, but on enabling instructors to move between expressive high-level interaction, precise low-level control, and context-aware grounding. Systems that foreground such multi-level oversight are better positioned to support responsible and pedagogically sound use of generative models.}

% \subsection{Student Simulation as an Exploratory Design Instrument}

% Another key insight from our study is that LLM-based student simulation can function as an exploratory, XAI-like mechanism that helps instructors reason about diverse student thinking patterns during assessment design. By exposing simulated high-frequency reasoning strategies and allowing instructors to interactively adjust student profile parameters (e.g., cognitive ability levels within each learner cluster), simulation enables instructors to explore how different types of students might approach the same question. This capability supports instructors in examining not only whether a question is answerable, but how it may elicit qualitatively different reasoning processes across learner populations.

% Feedback from the large-scale online study suggests that, at an aggregate level, LLM-simulated high-frequency reasoning strategies were often able to cover students’ self-reported thinking processes. Many participants were able to identify a simulated strategy that closely matched their own reasoning, indicating that the simulation captured common and instructional-relevant patterns of student thinking. This coverage supports the use of simulation as a means to surface dominant reasoning pathways that instructors may wish to encourage, diagnose, or challenge through question design.

% \textit{\textbf{Student Simulation in Assessment Design: Value, Limits, and Design Implications. }}

\subsection{\zixin{Student Simulation in Assessment Design: Value, Limits, and Design Implications. }}
\zixin{
% Rather than serving as a predictive model of individual learner behavior, our study indicates that the primary value of LLM-based student simulation lies in supporting instructors’ exploration of diverse reasoning patterns that may arise during assessment design. 
While we demonstrate that simulation can capture aggregate reasoning patterns, its primary pedagogical value lies in supporting instructors' exploration of diverse reasoning patterns that may arise during assessment design. 
By making such patterns explicit, simulation enables instructors to examine both the strengths and the limitations of their assumptions about how students reason through visualization literacy tasks.}

\zixin{Results from the large-scale online study indicate that, at an aggregate level, LLM-simulated high-frequency reasoning strategies were often able to cover students’ self-reported thinking processes. Many participants identified at least one simulated strategy that closely matched their own reasoning, suggesting that simulation can effectively surface dominant and instructionally relevant reasoning patterns shared by many learners. This form of aggregate-level coverage is particularly valuable for assessment design, where instructors often need to reason about common misconceptions, typical solution paths, and broadly shared difficulties rather than about individual cognitive trajectories.}

% At the same time, our findings suggest the presence of blind spots in modeling opportunistic and heuristic reasoning strategies that can arise in authentic student responses. While simulated students were relatively effective at capturing structured or “idealized” reasoning patterns—such as stepwise comparison or careful interpretation—they were less successful in reflecting minimally effortful or shortcut-driven behaviors. For example, some participants reported relying on rough visual estimation (e.g., ``I just eyeballed the bar lengths and chose the closest answer'') rather than the multi-step reasoning reflected in the simulated strategies. These cases point to a gap between modeled reasoning and the pragmatic, sometimes coarse strategies that students employ in practice.

\zixin{At the same time, our findings suggest the presence of blind spots in modeling opportunistic and heuristic reasoning strategies that can arise in authentic student responses. While simulated students were relatively effective at representing structured or “idealized” reasoning patterns—such as stepwise comparison or careful interpretation—they often failed to capture minimally effortful or shortcut-driven behaviors. For example, some participants reported relying on rough visual estimation (e.g., ``I just eyeballed the bar lengths and chose the closest answer'') rather than the multi-step reasoning reflected in the simulated strategies. These cases highlight a gap between modeled reasoning and the pragmatic, sometimes coarse strategies that students employ in practice.}

\zixin{These limitations point to important design implications for future student simulation systems. While our current student profiles—constructed primarily from demographic attributes, cognitive ability levels, and targeted knowledge points—were sufficient to surface many common reasoning patterns, they remain incomplete. Prior work in the learning sciences suggests that students’ task performance is also shaped by factors such as metacognitive regulation, motivation, affective state, and engagement, which are not well captured by ability- or knowledge-centric profiles alone. Incorporating such dimensions into future simulation approaches may enable richer and more realistic representations of student reasoning, though doing so will require closer grounding in learning theory and empirical studies of learner cognition.}

\zixin{Overall, our findings position LLM-based student simulation as a reflective design instrument rather than a ground-truth model of learner behavior. By making plausible reasoning pathways and their breakdowns visible, simulation supports instructors' exploration and comparison of how assessment designs may interact with learner diversity, while leaving pedagogical judgment and validation firmly in human hands.}

\subsection{\zixin{Implications for Visualization Literacy Education}}

\zixin{Our study offers several observations that inform broader discussions in visualization literacy education. In our sample, participants reported relatively high confidence in their visualization-related abilities, yet their baseline performance suggested more limited mastery of core concepts. In Stage~1, average self-assessed visualization ability exceeded four on a five-point scale, while the pre-test score averaged $2.11$ out of $5$. Although this result should be interpreted cautiously, it points to a mismatch between learners' perceived competence and their demonstrated ability on visualization literacy tasks.}

\zixin{This mismatch reflects a broader challenge in visualization education: difficulties often lie not in surface-level interpretation, but in reasoning about misleading encodings, proportional relationships, and contextual cues. Such gaps are unlikely to be revealed by static or correctness-only assessments. From this perspective, formative assessments that foreground reasoning, rather than answers alone, are particularly important for visualization literacy. Simulation-informed question design offers one way to prompt instructors to consider how different reasoning strategies, misunderstandings, or shortcuts may emerge, and to design questions that more effectively surface conceptual weaknesses.}

% \subsection{Ethical, Reproducibility, and Future Research Considerations}

% Despite the insights gained, our approach raises important ethical and reproducibility concerns. LLM-based simulation may encode biases related to demographic attributes or educational background, and evolving model versions complicate exact replication of results. While we mitigate these risks through de-identification, optional learner profiling, and explicit instructor control, future research should investigate bias-aware simulation audits, hybrid approaches that integrate empirical learner data, and standardized practices for documenting human–AI co-design decisions.

% Longitudinal classroom deployments are also needed to understand how instructors appropriate simulation over time and how simulation-informed design practices influence teaching strategies and learning outcomes in authentic educational contexts. Together, these directions point toward a broader research agenda focused on responsible, transparent, and pedagogically grounded use of generative models in education.

\subsection{\zixin{Limitations and Ethical Considerations}}

\zixin{Our study has several limitations that should be acknowledged. First, both question generation and student simulation rely on off-the-shelf LLMs, whose behavior is shaped by training data, prompting strategies, and ongoing model updates beyond our control. Consequently, generated questions, answers, and simulated reasoning traces cannot be assumed to be uniformly accurate, unbiased, or reproducible across time and deployment contexts. These characteristics require LLM outputs to be treated as provisional artifacts rather than authoritative representations of correctness or learner behavior.}

\zixin{Second, while the simulation was effective at surfacing common reasoning patterns at an aggregate level, it remains an incomplete representation of real student behavior. Opportunistic, heuristic, or minimally effortful strategies were not consistently captured, and learner profiles defined by demographic attributes, cognitive ability levels, and knowledge points necessarily simplify the richness of authentic learning contexts. These constraints limit the interpretability of simulated reasoning and reinforce the positioning of simulation as a design aid rather than a predictive model.}

\zixin{In this work, we sought to mitigate these limitations through de-identification, optional learner profiling, and explicit instructor control over both simulation parameters and assessment content. Nevertheless, these measures represent only partial safeguards. Future work is needed to develop more robust practices for bias auditing, transparency, and reproducibility in simulation-supported assessment design.}

\subsection{\zixin{Generalizability and Future Directions}}

\zixin{Beyond the visualization literacy context examined in this study, our findings suggest several directions in which simulation-informed assessment design may be explored further. One recurring insight concerns the use of simulated reasoning beyond question authoring. Rather than remaining solely a design-time tool, representative reasoning paths—both correct and incorrect—could be selectively reused as reflective feedback for learners, supporting comparison and metacognitive reflection rather than answer verification.}

\zixin{Instructors also discussed the value of simulation for adapting instructional materials to learners with different prior knowledge or disciplinary backgrounds. By exploring how plausible reasoning patterns shift across learner profiles, instructors could use simulation to adjust question framing, explanations, or scaffolding when preparing materials for different audiences or instructional contexts. Similarly, simulation was discussed as a way to reason about group-level dynamics, such as how differences in background knowledge might shape collaborative learning. Together, these directions point toward broader applications of simulation-informed design in educational settings where learning hinges on complex reasoning processes and instructor judgment.}

\section{Conclusion}

We introduced \textit{\systemName}, a visual analytics system that supports educators in designing and refining MCQs for visualization literacy education. By integrating MLLM-based question generation with simulated student reasoning, \zixin{we enables instructors to explore and iteratively refine question designs with respect to diverse learner profiles at design time. Through a multi-method evaluation, we examine the value and limitations of this approach and surface design insights on how visual analytics and MLLMs can be combined to support instructor-centered, flexible, and responsible assessment design in visualization literacy and related educational contexts.}

\bibliographystyle{IEEEtran}

\bibliography{template}

%--------------1
\begin{IEEEbiography}[{\includegraphics[width=1in,height=1.25in,clip,keepaspectratio]{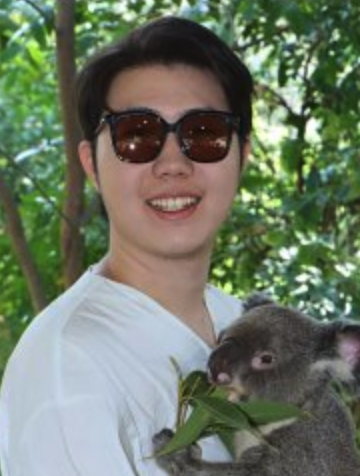}}]{Zixin Chen} is currently a Ph.D. candidate from Hong Kong University of Science and Technology. His research centers on Data Visualization and Human-AI Collaboration, with a particular emphasis on leveraging them in LLM for Education. He designs interactive tools that support teaching and pedagogical decision-making, personalize learning experiences, and makes LLM behavior more trustworthy for both learners and educators. In parallel, he explores the broader applicability of LLM-driven models and systems in other significant domains (AI4Health and AI4Science). More information: \url{https://cinderd.github.io/}.
\end{IEEEbiography}
%--------------2
\begin{IEEEbiography}[{\includegraphics[width=1in,height=1.25in,clip,keepaspectratio]{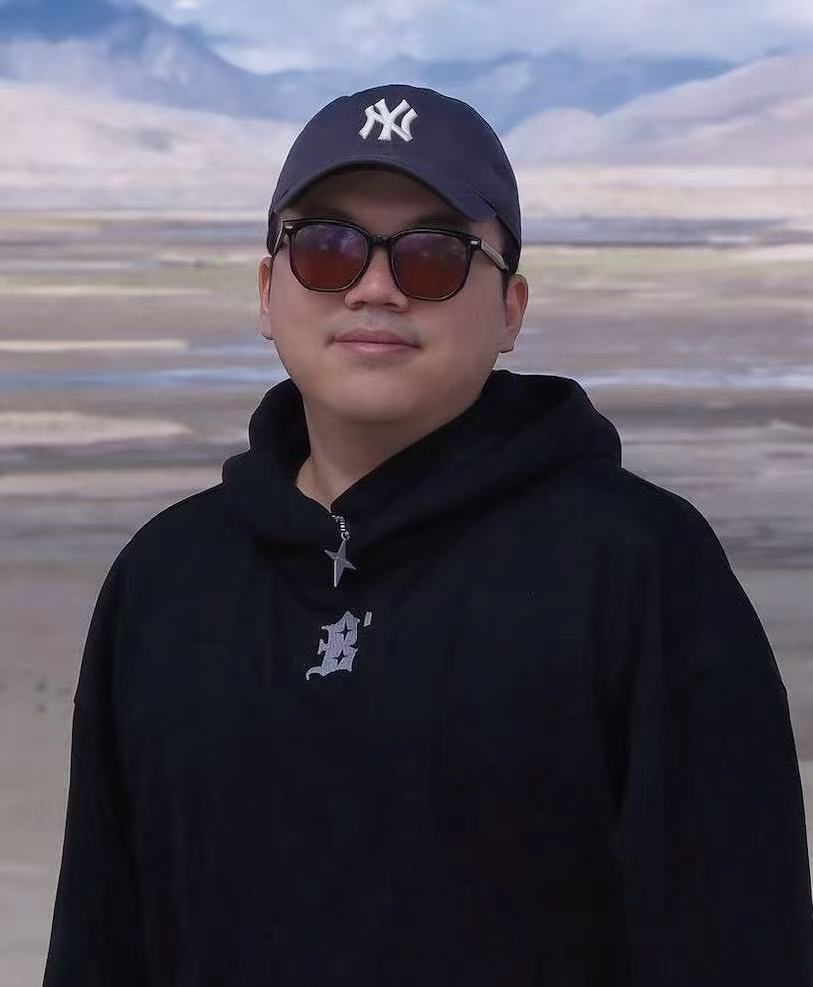}}]{Yuhang Zeng} is currently s a Master’s student in Computational Data Science at Carnegie Mellon University. He received his B.S. degree in Data Science and Technology and Computer Science from the Hong Kong University of Science and Technology. His research interests include data visualization, human–AI interaction, and interactive AI systems, with a particular focus on integrating large language models into visual analytics and educational tools. 
\end{IEEEbiography}
%--------------3
\begin{IEEEbiography}[{\includegraphics[width=1in,height=1.25in,clip,keepaspectratio]{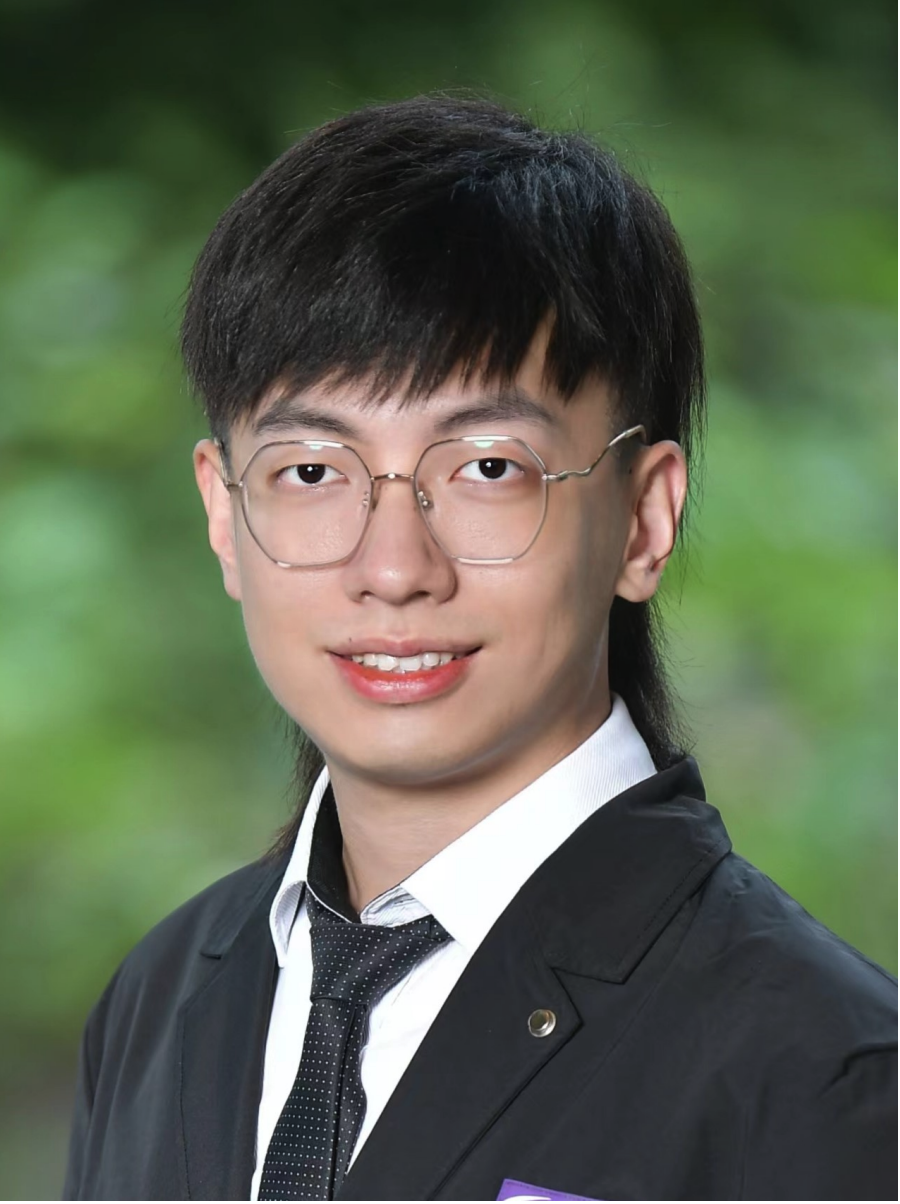}}]{Sicheng Song}
received Ph.D. from the School of Computer Science and Technology at East China Normal University, in 2024. He is a Postdoctoral Research Fellow in the Computer Science and Engineering at the Hong Kong University of Science and Technology. His main research interests include information visualization and visual analysis. For more details, please refer to \url{https://ByShawn.github.io/}.
\end{IEEEbiography}
%--------------4
\begin{IEEEbiography}[{\includegraphics[width=1in,height=1.25in,clip,keepaspectratio]{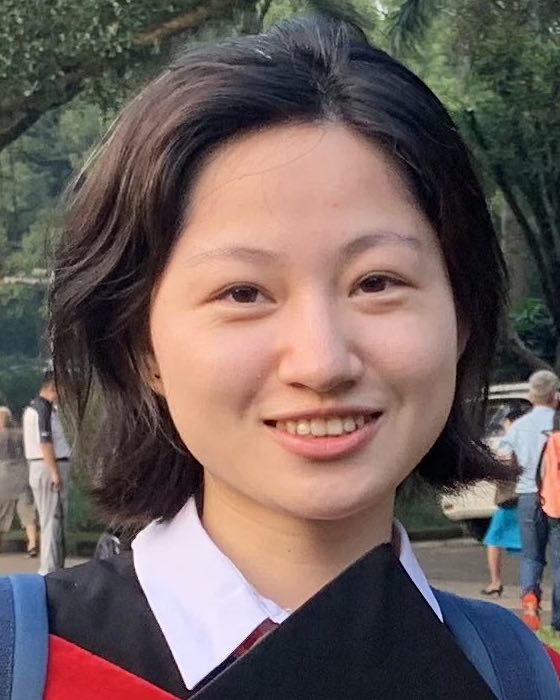}}]{Yanna Lin} is currently a postdoctoral researcher at the University of Waterloo.
She received her Ph.D. from the Hong Kong University of Science and Technology and her B.Eng. from Sun Yat-sen University.
Her research interests include data visualization and human–computer interaction (HCI), with a focus on designing intuitive interfaces and developing algorithms to enhance data communication and analysis, ultimately supporting data-driven decision-making. For more information, please visit: \url{https://yannahhh.github.io/}.
\end{IEEEbiography}
%--------------5
\begin{IEEEbiography}[{\includegraphics[width=1in,height=1.25in,clip,keepaspectratio]{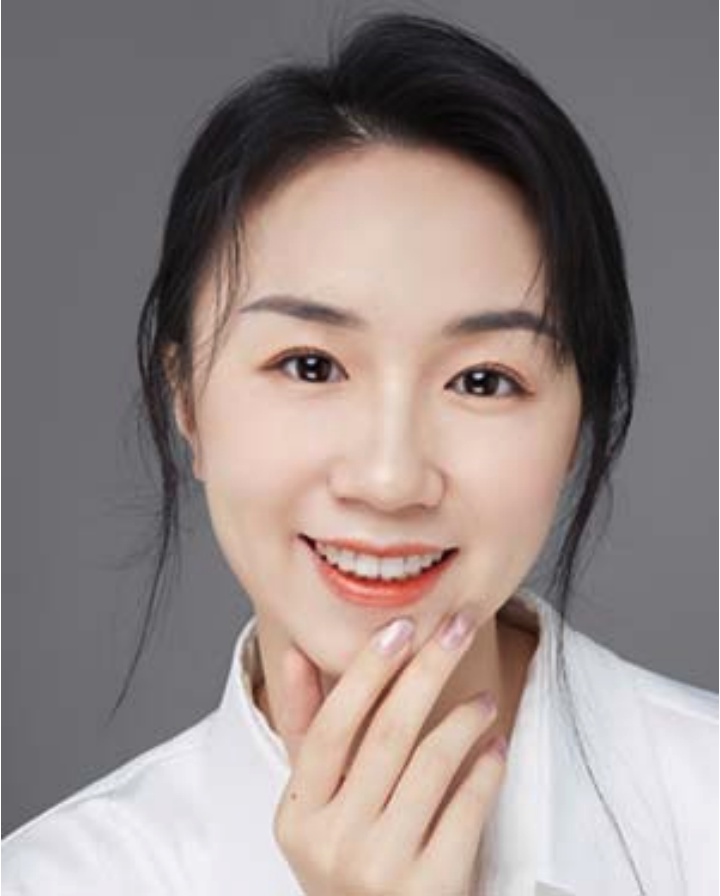}}]{Xian Xu} received the BA degree in directing from the
Central Academy of Drama, China, the MFA degree
in Studies of Drama, and the PhD degree in computational media and arts from HKUST VisLab. She
is a tenure-track Assistant Professor at the Department of Digital Arts and Creative Industries, Lingnan University, Hong Kong. She
is a visiting fellow at the University of Oxford and
the University of Cambridge. Apart from her research
work, she is also a television and film scriptwriter.
\end{IEEEbiography}
%--------------6
% \begin{IEEEbiography}[{\includegraphics[width=1in,height=1.25in,clip,keepaspectratio]{zhouyan.png}}]{Yan Zhou} received her bachelor’s degree in biology from the University of Manchester and Shandong University, followed by a master’s degree in computational biology from University College London. She joined Prof. Tiannan Guo's lab as a PhD student in 2021, with research focusing primarily on clinical proteomicså.
% \end{IEEEbiography}
% %--------------7
% \begin{IEEEbiography}[{\includegraphics[width=1in,height=1.25in,clip,keepaspectratio]{ruanshaolun.png}}]{Shaolun Ruan} is a Ph.D. candidate of Computer Science at Singapore Management University, under the supervision of Professor Yong WANG and Professor Jiannan LI. His research interests include Data Visualization and Human-Computer Interaction. His work focuses on developing human-centered computing tools to address complex scientific problems, facilitating the process of explainability and data-driven decision-making. More information: \url{https://shaolun-ruan.com/}.
% \end{IEEEbiography}
%--------------8
\begin{IEEEbiography}[{\includegraphics[width=1in,height=1.25in,clip,keepaspectratio]{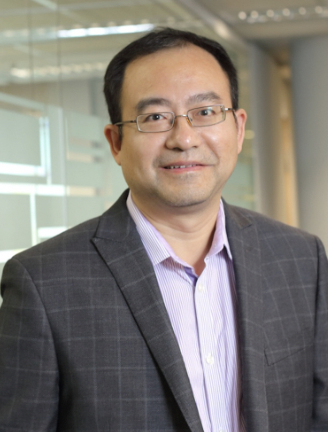}}]{Huamin Qu} is the dean of the Academy of Interdisciplinary Studies, head of the Division of Emerging Interdisciplinary Areas, and a chair professor in the Department of Computer Science and Engineering at HKUST. He obtained a BS in Mathematics from Xi’an Jiaotong University, China, an MS, and a PhD in Computer Science from the Stony Brook University. His main research interests are in visualization and human-computer interaction, with focuses on urban informatics, social network analysis, Elearning, text visualization, and explainable artificial intelligence More information: \url{http://huamin.org/}.
\end{IEEEbiography}

%--------------5
\begin{IEEEbiography}[{\includegraphics[width=1in,height=1.25in,clip,keepaspectratio]{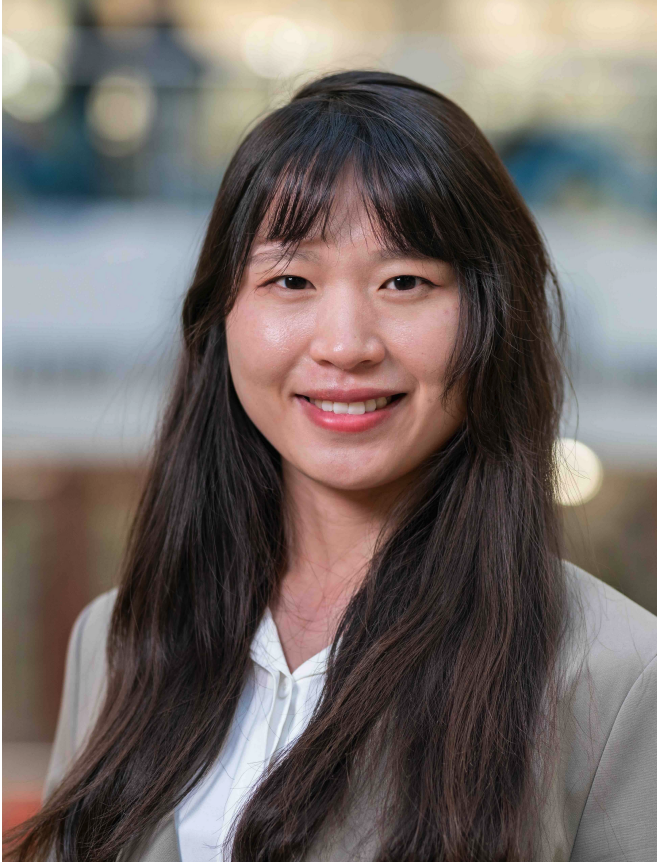}}]{Meng Xia} is an assistant professor in the Department of Computer Science and Engineering at
Texas A\&M University. She obtained a Ph.D. degree
from the Hong Kong University of Science and
Technology and worked as a Postdoc at Carnegie
Mellon University and Korea Advanced Institute of
Science and Technology, respectively. Her research
interests mainly focus on Human-AI Interaction,
Data Visualization, and Education Technology. More
details can be found at https://www.xiameng.org.
\end{IEEEbiography}

% \end{thebibliography}

% \newpage

% \section{Biography Section}
% If you have an EPS/PDF photo (graphicx package needed), extra braces are
%  needed around the contents of the optional argument to biography to prevent
%  the LaTeX parser from getting confused when it sees the complicated
%  $\backslash${\tt{includegraphics}} command within an optional argument. (You can create
%  your own custom macro containing the $\backslash${\tt{includegraphics}} command to make things
%  simpler here.)
 
% \vspace{11pt}

% \bf{If you include a photo:}\vspace{-33pt}
% \begin{IEEEbiography}[{\includegraphics[width=1in,height=1.25in,clip,keepaspectratio]{fig1}}]{Michael Shell}
% Use $\backslash${\tt{begin\{IEEEbiography\}}} and then for the 1st argument use $\backslash${\tt{includegraphics}} to declare and link the author photo.
% Use the author name as the 3rd argument followed by the biography text.
% \end{IEEEbiography}

% \vspace{11pt}

% \bf{If you will not include a photo:}\vspace{-33pt}
% \begin{IEEEbiographynophoto}{John Doe}
% Use $\backslash${\tt{begin\{IEEEbiographynophoto\}}} and the author name as the argument followed by the biography text.
% \end{IEEEbiographynophoto}

\vfill

\end{document}